\documentclass[aps,reprint,superscriptaddress]{revtex4-2}
\usepackage[utf8]{inputenc}
\usepackage{physunits}
\usepackage{amsmath}
\usepackage{hyperref}
\usepackage{amssymb}
\usepackage{physics}
\usepackage{float}
\usepackage{graphicx}
\usepackage{siunitx}
\usepackage{bbm}
\usepackage{appendix}
\usepackage[table,xcdraw]{xcolor}
\usepackage{mathrsfs}
\usepackage{circuitikz}
\usepackage{listings}
\usepackage{xcolor}
\usepackage{siunitx}
\usepackage{dcolumn}
\usepackage{bm}
\usepackage{cleveref}
\usepackage{comment}
\usepackage{soul}
\DeclareMathOperator\arctanh{arctanh}

\newcommand{\wc}{\omega_{\mathrm{cav}}}
\newcommand{\wq}{\omega_{\mathrm{q}}}
\newcommand{\wm}{\Omega_{\mathrm{m}}}
\newcommand{\wdr}{\omega_{\mathrm{d}}}
\newcommand{\Deltac}{\Delta_{\mathrm{c}}}
\newcommand{\Deltaq}{\Delta_{\mathrm{q}}}
\newcommand{\Ac}{\mathcal{E}_{\mathrm{c}}}
\newcommand{\ncav}{n_{\mathrm{cav}}}
\newcommand{\ncrit}{n_{\mathrm{crit}}}

\newcommand{\ba}{\hat{a}}
\newcommand{\bb}{\hat{b}}

\newcommand{\bs}{\hat{\sigma}}

\crefname{pluralequation}{eqs.}{Eqs.}
\begin{document}

\title{Strongly driven cavity quantum electrodynamical-optomechanical hybrid system}

\author{Xuxin Wang}
\email[Correspondence author: ]{xuxin.wang@epfl.ch}
\author{Jiahe Pan}
\author{Tobias J. Kippenberg}
\author{Shingo Kono}

\email[Correspondence author: ]{shingo.kono@nbi.ku.dk}
\thanks{\\
Current address: Niels Bohr Institute, University of Copenhagen}
\affiliation{%
 Laboratory of Photonics and Quantum Measurements (LPQM),
Swiss Federal Institute of Technology Lausanne (EPFL), Lausanne, Switzerland
}%

\affiliation{Center for Quantum Science and Engineering, EPFL, Lausanne, Switzerland}

\begin{abstract}

Hybrid quantum systems harness the distinct advantages of different physical platforms, yet their integration is not always trivial due to potential incompatibilities in operational principles. Here, we theoretically propose and demonstrate a scheme for generating non-Gaussian mechanical states using a strongly driven hybrid system that combines cavity quantum electrodynamics (QED) and cavity optomechanics. Our protocol prepares a non-Gaussian cavity state in the dispersive regime of cavity QED and subsequently transfers it to a mechanical oscillator using the optomechanical interaction enhanced by a coherent cavity drive. While non-Gaussian cavity state control in cavity QED is well established in the dispersive regime, its behavior under strong cavity drive, essential for cavity optomechanics, remains largely unexplored. To bridge this gap, we develop an efficient simulation framework to model cavity QED dynamics in the high-photon-number regime. We show that a strong cavity drive can coherently displace the cavity state with minimal perturbations, effectively decoupling it from the qubit. The resulting large coherent cavity field enhances the optomechanical coupling strength, enabling high-fidelity transfer of non-Gaussian cavity states to the mechanical mode. These results reveal new dynamical features of driven cavity QED and open a pathway toward realizing non-Gaussian mechanical quantum memories and sensors.
\end{abstract}

\maketitle

\section{Introduction}

\begin{figure}[htp!]
    \centering
    \includegraphics[width = 8.5cm]{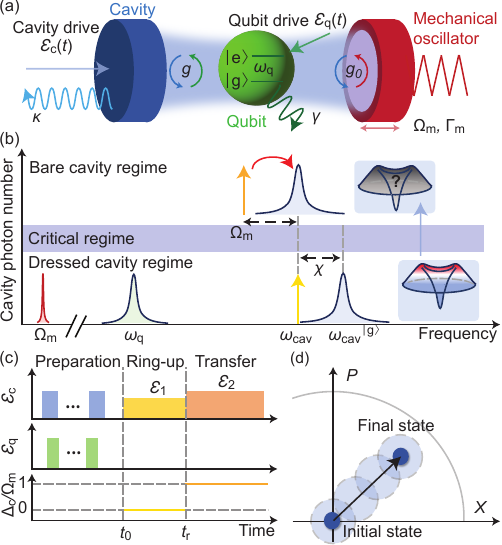}
    \caption{%
        (a)~Schematic of the cavity QED-optomechanical hybrid system. The qubit (green) is coupled to a cavity field (blue) to construct a cavity QED system. One mirror of the cavity is vibrating, acting as a mechanical oscillator (red) that is coupled to the cavity field to construct a cavity optomechanical system. The cavity and the qubit can be individually controlled through the cavity drive and the qubit drive, respectively. 
        (b)~Frequency landscape of the cavity QED-optomechanical hybrid system. The non-Gaussian cavity state is generated in the dressed regime and successively transferred to the bare cavity regime by a strong cavity drive. %
        (c)~Drive scheme of our protocol. During the preparation step, a sequence of pulses are applied to both the qubit and the cavity to generate an arbitrary non-Gaussian cavity state. To transfer the state to the mechanical oscillator, the cavity is first driven to a sufficiently high photon number, thereby enhancing the optomechanical coupling rate. In the transfer step, the cavity coherent field is in a forced oscillation at the red-detuned cavity frequency, which triggers the stationary optomechanical interaction and transfers the cavity state to the mechanical oscillator. 
        (d)~ Simulation method. The system is simulated in the time-dependent displaced frame for each bosonic mode, which are adaptively updated to minimize the required Hilbert space, as indicated by the blue circles with dashed outline. 
        }
    \label{FIG-1} 
\end{figure}

%


Cavity optomechanics based on superconducting circuits (circuit optomechanics) has emerged as a powerful platform for controlling mechanical oscillators at the quantum level \cite{Aspelmeyer2014}. Integrating high-$Q$ mechanical membranes with superconducting microwave circuits has enabled cavity optomechanics in the microwave domain and has already facilitated a broad range of quantum protocols, including ground-state cooling~\cite{Teufel2011s}, coherent coupling~\cite{Teufel2011c}, mechanical squeezing~\cite{Lecocq2015q, Wollman2015, Youssefi2022}, entanglement generation~\cite{Palomaki2013e, OckeloenKorppi2018, Kotler2021}, and quantum transduction~\cite{Andrews2014}.
Recent advances in circuit optomechanics also offer key advantages, such as long coherence times~\cite{Youssefi2023} and scalability~\cite{Youssefi2022, Chegnizadeh2024}, making the platform particularly promising for applications in quantum memory and sensing \cite{Youssefi2023, Seis2022, Hertzberg2009}. 
However, a key limitation of the cavity optomechanical systems is their reliance on strong electromagnetic drives to enhance intrinsically weak radiation-pressure-type interactions. This enhancement results from the \textit{linearization} process, reducing the interaction to an effective bilinear form \cite{Aspelmeyer2014}. Therefore, it restricts access to non-Gaussian operations, which are essential for universal quantum control~\cite{Lloyd1999}.

Cavity quantum electrodynamics (QED), particularly its superconducting implementation, known as circuit QED~\cite{Blais2021}, has enabled the generation and manipulation of non-Gaussian quantum states of superconducting circuits by leveraging the nonlinearity of Josephson junctions. Importantly, the framework has been extended to quantum acoustics, i.e., hybrid quantum systems in which superconducting qubits are coherently coupled to gigahertz-frequency acoustic modes through piezoelectric effects~\cite{o2010}. While these developments have enabled advanced non-Gaussian quantum control of mechanical systems~\cite{Chu2018, Satzinger2018, wollack2022, Bild2023, Yang2024}, it remains a significant challenge to extend such control to megahertz-frequency mechanical membranes within the framework of cavity optomechanics~\cite{Pirkkalainen2015, Lecocq2015, Ma2021, Palomaki2013c}.

In this work, we theoretically propose a novel scheme for generating non-Gaussian mechanical membrane states in a hybrid system that seamlessly combines cavity optomechanics with cavity QED.
A fundamental challenge arises from the different operating regimes of these two systems with respect to cavity photon number. Cavity optomechanics generally demands a large cavity photon number to enhance the intrinsically weak optomechanical coupling. In contrast, cavity QED is typically operated in the low-photon-number regime, where a wide range of non-Gaussian cavity-state generations has been demonstrated~\cite{Vlastakis2013, Bretheau2015, Heeres2017, CampagneIbarcq2020, Kudra2022}.
For realizing the target hybrid quantum system, it is therefore essential to understand cavity QED dynamics in the high-photon-number regime. Even in different contexts, strongly driven cavity QED has recently attracted significant interest due to its relevance in attempts to realize high-fidelity qubit readout and gate operations with minimal backaction from strong drives. This has led to the development of advanced readout protocols~\cite{Reagor2018, Dassonneville2020, Sunada2024} and qubit cloaking techniques~\cite{Lledo2023, MunozArias2023}, as well as studies of measurement-induced state leakage~\cite{Sank2016} and transmon ionization~\cite{Dumas2024}. Despite these advances, a comprehensive theoretical framework for cavity QED in the high-photon-number regime remains largely undeveloped.

In our approach, we exploit two distinct linear-cavity-response regimes of the cavity QED system under off-resonant coupling: the dressed- and bare-cavity regimes. These two regimes are parameterized by the cavity photon number and separated by the critical photon number, which defines the crossover boundary between them~\cite{Blais2021}. 
More specifically, we propose using the dressed cavity regime to synthesize a non-Gaussian cavity state, then switching to the bare cavity regime with a strong coherent cavity drive to enhance the optomechanical interaction, enabling the non-Gaussian cavity state to be faithfully transferred to the mechanical oscillator.
To assess the feasibility of this strategy, we identify two essential questions that must be resolved. First, can the cavity quantum state be coherently displaced without introducing unwanted perturbations as the mean photon number is pushed beyond the critical photon number? Second, once the cavity state is displaced to the bare cavity regime, can it be fully decoupled from qubit dynamics, such that the system evolves under pure optomechanical interaction alone, with negligible residual qubit back-action?

To address these questions, we study the dynamics of a strongly driven hybrid quantum system of cavity QED and cavity optomechanics. Building on the insight of prior work~\cite{Goto2023}, we adopt a novel method to efficiently simulate the dynamics of the hybrid system. By adaptively shifting the displacement frame and tracking the trajectory of coherent amplitudes of the cavity and the mechanical oscillator, we can truncate the Hilbert space to enhance the computational efficiency. 
Within the framework of our method, we confirm that a non-Gaussian mechanical state can be generated using the experimentally achievable parameters.
Furthermore, by adopting perturbation theory, we analytically predict that the dominant cavity-state deformations are phase shifting and squeezing, which can be significantly suppressed by increasing the coherent cavity photon number.


%
\section{Model and simulation method}
\label{Sec:model}

In Fig.~\ref{FIG-1}a, we illustrate the schematic of the cavity QED-optomechanical hybrid system. A cavity is coupled to an atom (qubit) to construct a cavity QED system. The qubit is a two-level system with two energy eigenstates: the ground state $\ket{g}$ and the excited state $\ket{e}$. Furthermore, the cavity is coupled to a mechanical oscillator to construct a cavity optomechanical system. The Hamiltonian of the hybrid system is given by 
\begin{equation}
\begin{aligned}
        \hat H_\mathrm{s}/\hbar = &\wc\hat a^{\dag}\hat a+\wq\dfrac{\hat\sigma_z}{2} +\wm\hat b^{\dag}\hat b \\
        &+  g(\ba^{\dag}\bs_-+\ba\bs_+)+ g_0\ba^{\dag}\ba(\bb^{\dag}+\bb),
\end{aligned}
\end{equation}
with $\hat a$ ($\hat a^{\dag}$) the annihilation (creation) operator of the cavity mode, $\hat b$ ($\hat b^{\dag}$) the annihilation (creation) operator of the mechanical mode, $\hat \sigma_-$ ($\hat\sigma_+$) the Pauli annihilation (creation) operator of the qubit, $\wc$ the cavity frequency, $\wq$ the qubit frequency, $\wm$ the frequency of the mechanical oscillator, $g$ the coupling rate between the qubit and the cavity, and $g_0$ the single-photon optomechanical coupling rate. The coupling between the qubit and the cavity is approximated by the beam-splitter interaction, as the counter-rotating terms in the dipole-dipole interaction are neglected by the rotating-wave approximation (RWA)~\cite{Blais2021}. 

In our scheme, the main objective is to generate a non-Gaussian state in the mechanical oscillator. The whole procedure consists of three steps, as illustrated by the corresponding frequency landscape and drive scheme in Figs.~\ref{FIG-1}b and c. In the preparation step, a non-Gaussian state is prepared in the cavity by applying a sequence of pulses to the cavity and qubit. Since protocols for generating non-Gaussian cavity states in a cavity-QED setup are well-established~\cite{Kirchmair2013, Hofheinz2008, Hofheinz2009, CampagneIbarcq2020, Vlastakis2013, Holland2015}, we will not include the simulation for the preparation step. All the simulations presented in this work focus on the subsequent dynamics, beginning after the state preparation.
In the following ring-up step, a strong cavity drive is applied to increase the coherent cavity photon number and displace the non-Gaussian state in the cavity. The cavity drive Hamiltonian is expressed as
\begin{equation}
    \hat H_{\text d}(t)/\hbar= \dfrac{\Ac(t)}{2}\hat a^{\dag}e^{-i\wdr t}+\dfrac{\Ac^*(t)}{2}\hat a e^{i\wdr t},
\end{equation}
where $\wdr$ is the drive frequency, which matches the bare cavity frequency $\wc$ to efficiently increase the cavity photon number. $\Ac(t)$ is the complex drive amplitude. The transfer step begins once the cavity photon number becomes sufficiently large to achieve the desired optomechanical coupling rate. In order to retain a stationary optomechanical coupling while transferring the cavity state to the mechanical oscillator, the cavity coherent amplitude should be in forced oscillation at the lower sideband of the cavity $\wc-\wm$. The cavity drive is shifted non-adiabatically by the mechanical frequency, and the amplitude and phase of the drive are properly configured such that the cavity coherent amplitude before switching matches that of the corresponding forced oscillation, which avoids possible transient cavity dynamics after entering the transfer step. Therefore, the cavity quantum state can be transferred to the mechanical oscillator via the optomechanical coupling, enhanced by the stationary cavity field in the forced oscillation.

Based on the idea of our protocol, it is important to understand how the non-Gaussian state evolves during the entire transfer step. However, the analytical solution of the driven JC Hamiltonian has not been found if the drive amplitude is greater than the coupling rate between the cavity and the qubit \cite{Alsing1992}. This encourages us to apply an efficient method to numerically simulate the dynamics of the strongly driven system. Due to the large cavity photon number induced by a strong cavity drive, it is difficult to perform a full simulation with a Hilbert space that captures the actual photon number. Therefore, we develop a novel simulation method based on the displaced frame that is adaptively updated to follow the trajectory of coherent amplitudes for the cavity and the mechanical oscillator, as illustrated in Fig.~\ref{FIG-1}d. This reduces the computational overhead without undermining accuracy.

In the numerical simulation, we first go to the rotating frame of the drive field frequency for both the cavity and qubit, eliminating all time-dependent terms in the Hamiltonian of the driven hybrid system. The total Hamiltonian is expressed as
\begin{equation}
\begin{aligned}
    \hat{H}/\hbar 
    =&\, \Deltac \hat{a}^\dagger \hat{a} +  \Deltaq \frac{\hat{\sigma}_z}{2} +  \wm \hat{b}^\dagger \hat{b} \\
    &+  g (\hat{a}^\dagger \hat{\sigma}_- + \hat{a} \hat{\sigma}_+) +  g_0 \hat{a}^\dagger \hat{a} (\hat{b}^\dagger + \hat{b}) \\
    &+ \frac{ \Ac}{2} \hat{a}^\dagger + \frac{ \Ac^*}{2} \hat{a},
\end{aligned}
\label{eq:total_H}
\end{equation}
where $\Deltac = \wc - \wdr$ and $\Deltaq = \wq - \wdr$ denote the detunings of the cavity and qubit frequencies from the drive frequency, respectively.
The initial state is denoted by $\hat{\rho}_0$, and the total simulation time spans from $t = 0$ to $t = T$. We divide this total duration into $N \in \mathbb{N}$ segments with the interval of $\tau = T/N$.

In the $n$-th interval ($n=0,1,\cdots,N-1$), the simulation starts from an initial state $\hat{\rho}_n(0)$, obtained by displacing the final state $\hat{\rho}_{n-1}(\tau)$ of the previous step to the origins of the cavity and mechanical phase spaces, i.e.
\begin{equation}
    \hat{\rho}_n(0) = \hat{D}^\dagger(\delta\alpha_{n}, \delta\beta_{n}) \hat{\rho}_{n-1}(\tau) \hat{D}(\delta\alpha_{n}, \delta\beta_{n}),
\end{equation}
where the optimal displacement amplitudes are found to be
\begin{align}
    \delta\alpha_{n} &= \text{Tr}[\hat{a} \hat{\rho}_{n-1}(\tau)], \label{eq:da}\\
    \delta\beta_{n} &= \text{Tr}[\hat{b} \hat{\rho}_{n-1}(\tau)]. \label{eq:db}
\end{align}
These displacements ensure that $\text{Tr}[\hat{a} \hat{\rho}_n(0)] = 0$ and $\text{Tr}[\hat{b} \hat{\rho}_n(0)] = 0$ for all $n$, effectively reducing the size of the required Hilbert space for each interval.
Here, the combined displacement operator is defined as $\hat{D}(\alpha, \beta) = \hat{D}_{\hat{a}}(\alpha)\hat{D}_{\hat{b}}(\beta)$, where 
$\hat{D}_{\hat{a}}(\alpha) = \exp(\alpha \hat{a}^\dagger - \alpha^* \hat{a})$ and 
$\hat{D}_{\hat{b}}(\beta) = \exp(\beta \hat{b}^\dagger - \beta^* \hat{b})$ are the displacement operators for the cavity and the mechanical mode, respectively.
Note that the initial state for $n = 0$ is obtained from $\hat{\rho}_{-1}(\tau) \equiv \hat{\rho}_0$.

Accordingly, the master equation governing the dynamics in the $n$-th interval ($t \in [0, \tau]$) is modified as
\begin{equation}
    \frac{\mathrm{d}\hat{\rho}_n}{\mathrm{d}t} = -\frac{i}{\hbar}[\hat{H}^{\text{D}}_n, \hat{\rho}_n] 
    + \kappa \mathcal{D}[\hat{a} + \alpha_n] \hat{\rho}_n 
    + \gamma \mathcal{D}[\hat{\sigma}_-] \hat{\rho}_n,
    \label{eq:master_equation}
\end{equation}
where the Hamiltonian is modified in accordance with the displaced frames, as
\begin{equation}
    \hat{H}^{\text{D}}_n = \hat{D}^\dagger(\alpha_n, \beta_n) \hat{H} \hat{D}(\alpha_n, \beta_n),
\end{equation}
and the accumulated displacement amplitudes up to the $n$-th interval can be obtained as
\begin{align}
    \alpha_n &= \sum_{i=0}^{n} \delta\alpha_i, \\
    \beta_n  &= \sum_{i=0}^{n} \delta\beta_i.
\end{align}

Furthermore, the cavity Lindblad dissipator is also displaced, and takes the form $\mathcal{D}[\hat{a} + \alpha_n]$, where the standard dissipator is defined as
\begin{equation}
    \mathcal{D}[\hat{a}] \hat{\rho} = \hat{a} \hat{\rho} \hat{a}^\dagger - \frac{1}{2}\left\{ \hat{a}^\dagger \hat{a}, \hat{\rho} \right\}.
\end{equation}
Note that the mechanical dissipation is omitted in our simulation, as discussed later.

By iteratively solving the master equation while updating the displacement frames step by step, the final quantum state in the original frame at time $T$, i.e., without displacement frame transformations, is obtained by correcting for the accumulated displacements as
\begin{equation}
    \hat{\rho}_\mathrm{sim}(T) = \hat{D}(\alpha_{N}, \beta_{N}) \hat{\rho}_\mathrm{sim}^D(T) \hat{D}^\dagger(\alpha_{N}, \beta_{N}), 
    \label{eq:center_amp}
\end{equation}
where $\hat{\rho}_\mathrm{sim}^D(T) = \hat{\rho}_{N}(0)$ is the quantum state centered around the coherent amplitudes $\alpha_{N}$ and $\beta_{N}$ at time $T$, representing the output of our numerical simulation. 
This state satisfies $\mathrm{Tr}[\hat{a} \hat{\rho}_\mathrm{sim}^D(T)] = \mathrm{Tr}[\hat{b} \hat{\rho}_\mathrm{sim}^D(T)] = 0$. 
Moreover, the coherent photon and phonon numbers in the original frame are given by $|\alpha_N|^2$ and $|\beta_N|^2$, respectively.
In this way, we can reconstruct the full quantum dynamics of the system with reduced computational overheads.

We benchmark our method by performing analytically solvable dynamics of a coherently driven cavity and show that the simulation errors can be sufficiently suppressed by properly choosing the size of the Hilbert space to be $N_{\mathrm{cav}}$ = 20 and the duration for each interval to be $\tau = 1$ ns for the parameter regimes considered in our work (see Appendix~\ref{App:benchmark}). All the simulations are performed with The Quantum Toolbox in Python (QuTip) \cite{Johansson2012}. 

In the simulation, we choose the parameters that have been realized in the state-of-the-art superconducting optomechanical circuits: $g_0/2\pi = 100$ Hz, $\wm/2\pi= 1$ MHz \cite{Teufel2011s, Youssefi2023}, while $\Delta/2\pi = (\wc -\wq )/2\pi = 100$ MHz and $g/2\pi = 5$ MHz are used and achieved by conventional circuit QED systems. 
The effects of the cavity decay rate $\kappa$ and the qubit decay rate $\gamma$ are studied by parameter sweep (see Sec.~\ref{Sec:transfer}). 
The critical photon number is given by $n_{\text{crit}} = \Delta^2/4g^2 = 100$. We aim to achieve a cavity photon number $n_{\text{cav}}\approx 10^6$ to realize a sufficiently large optomechanical coupling rate of $g_0\sqrt{n_{\text{cav}}}/2\pi\approx 100$ kHz. 
The mechanical thermal decoherence rate $\gamma_{\mathrm{th}} = \Gamma_{\text m} n_{\text{th}}$ is typically smaller than $\kappa$ and $\gamma$ by several orders of magnitude and thus is neglected in our numerical simulation, where $\Gamma_{\text m}$ is the bare mechanical dissipation rate and $n_{\text{th}}$ is the thermal bath occupation \cite{Youssefi2022}.



 \begin{figure*}[tp!]
    \centering
    \includegraphics[width = 17cm]{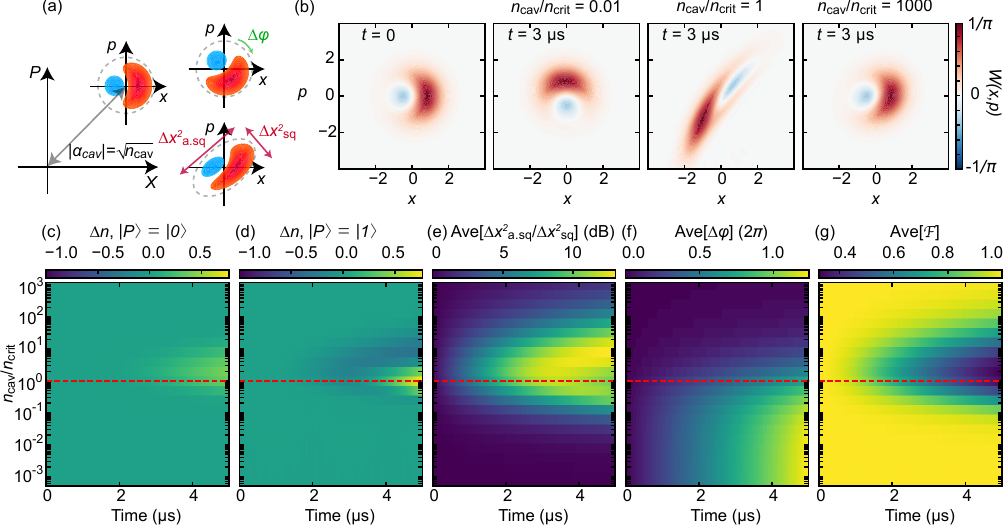}
    \caption{
    (a) Cavity-state dynamics under a displaced JC Hamiltonian and illustration of the leading effects of cavity-state deformations: phase shift and squeezing.
    (b) Wigner distribution of the displaced cavity states after $t = 3~\mu\text{s}$ of evolution, shown for different initial coherent cavity photon numbers. 
    (c,d) Photon number change $\Delta n$ for $\ket{P} = \ket{0}$ and $\ket{P} = \ket{1}$ as functions of time and the initial coherent photon number, respectively. 
    (e) Squeezing ratio $\Delta \hat{x}^2_{\text{a.sq}} / \Delta \hat{x}^2_{\text{sq}}$, averaged for $\ket{P} \in \{\ket{0}, \ket{1}\}$. 
    (f) Phase shift $\Delta\varphi$, averaged over $\ket{P} \in \{\ket{\pm}, \ket{\pm i}\}$. 
    (g) Fidelity of the cavity state to the initial state, averaged over $\ket{P} \in \mathcal{S}_6$. 
    The red dashed line indicates the boundary $n_{\mathrm{cav}} / n_{\text{crit}} = 1$.
    In the simulations, the cavity and the qubit are considered lossless, i.e., $\kappa/2\pi = \gamma/2\pi = 0$.
    }
    \label{FIG-2}
\end{figure*}

\section{Cavity dynamics in strongly driven cavity QED}
\label{Sec:non-gaussian}

We aim to understand the evolution of a non-Gaussian state throughout the entire protocol, especially how the cavity quantum state evolves when the coherent cavity photon number is close to and well above the critical photon number during the ring-up and transfer steps. For this purpose, in the following discussion, we will introduce three different models corresponding to variations of JC Hamiltonian. Besides, we will temporarily exclude the mechanical oscillator and focus only on the dynamics of the cavity QED system in different photon number regimes. 

\subsection{JC model in a cavity displaced frame}
\label{Subsec: non-driven JC}
To gain insight into the dynamics of the driven JC system within our protocol, we first study the JC Hamiltonian in a coherently displaced frame of the cavity without any drive, corresponding to the dynamics of a displaced cavity state in the standard JC model. This simplified model captures the leading-order deformations of the cavity state, which quantitatively define the boundaries between different photon-number regimes.

We start from the JC Hamiltonian in the rotating frame of the cavity frequency, which is expressed  as
\begin{equation}
    \hat H_{\text{JC}}/\hbar = \Delta\dfrac{\hat\sigma_z}{2} + g(\ba^{\dag}\bs_-+\ba\bs_+),
    \label{eq:H_JC}
\end{equation}
where $\Delta = \wq-\wc$ is the frequency difference between the cavity and the qubit. 
As schematically shown in Fig.~\ref{FIG-2}a, we let the initial state be a product state with the form of
\begin{equation}
    \ket{\Psi} = \hat D_{\hat a}(\alpha_{\mathrm{cav}})\ket{P}\otimes \ket{Q}.
    \label{eq:rho_target}
\end{equation}
The cavity is initialized in a certain state $\ket{P}$ that is displaced by $\alpha_{\mathrm{cav}}$, where the initial coherent cavity photon number is given by  $n_{\mathrm{cav}}=\abs{\alpha_{\mathrm{cav}}}^2$. To characterize the deformation of the displaced cavity state, we simulate the system dynamics using various initial cavity states $\ket{P}$ and different coherent photon numbers $n_{\mathrm{cav}}$, spanning values both below and above the critical photon number. Note that the qubit needs to be prepared in a specific state $\ket{Q}$ to mimic our actual protocol, which will be explained below.

In line with our numerical method mentioned in Sec.~\ref{Sec:model}, we simulate the cavity dynamics in a frame displaced by $\alpha_{\mathrm{cav}}$. Therefore, in the displaced frame, the Hamiltonian becomes
\begin{equation}
    \begin{aligned}
            \hat H_{\text {JC}}^{\text D}/\hbar =& \hat D^{\dag}_{\hat a}(\alpha_{\mathrm{cav}})\hat H_{\text{JC}}\hat D_{\hat a} (\alpha_{\mathrm{cav}})/\hbar\\
            =&  \Delta \dfrac{\hat\sigma_z}{2} + g(\hat a^{\dag}\hat\sigma_-+\hat a\hat\sigma_+) +  g\alpha_{\mathrm{cav}}\hat\sigma_x.
    \end{aligned}
    \label{eq:H_JC_D}
\end{equation}
Here, we assume $\alpha_{\mathrm{cav}}$ to be a real number for simplicity. We also assume $\omega_{\mathrm{q}}<\omega_{\mathrm{cav}}$, i.e. $\Delta<0$. 


Furthermore, the Hamiltonian of the qubit subsystem, including the effective qubit drive term from the cavity coherent amplitude, can be diagonalized using
\begin{equation}
    \hat R_y\left(\dfrac{\theta}{2}\right) = \exp(-i\dfrac{\theta}{2}\hat\sigma_y),
    \label{eq:Ry}
\end{equation}
where $\theta = \arctan(2g\alpha_{\mathrm{cav}}/\Delta)$.
As a result, the Hamiltonian is transformed to
\begin{equation}
    \begin{aligned}
        \hat H_{\mathrm{JC}}^{\mathrm{RD}}/\hbar = &\hat R_y^{\dag}({\theta}/{2})\hat H^{\text D}_{\text{JC}}\hat R_y({\theta}/{2})/\hbar\\
        =& -\tilde\Delta\dfrac{\hat\sigma_z}{2} + g_z(\hat a^{\dag}+\hat a)\dfrac{\hat\sigma_z}{2}\\
        &+  g_1\left(\hat a^{\dag}\hat\sigma_-+\hat a\hat\sigma_+\right)+  g_2\left(\hat a^{\dag}\hat\sigma_++\hat a\hat\sigma_-\right),
    \end{aligned}
    \label{eq:H_JC_RD_main}
\end{equation}
where
\begin{align}
    \tilde\Delta = & \sqrt{\Delta^2 + 4g^2\abs{\alpha_{\mathrm{cav}}}^2}, \\
    g_z =& g\sin\theta \label{eq:g_z},\\
    g_1 =& \dfrac{g}{2}(1+\cos\theta), \\
    g_2 =& \dfrac{g}{2}(1-\cos\theta). 
\end{align}

In our actual protocol, the qubit is initialized to the ground state before the cavity drive for the optomechanical coupling is applied.
Importantly, the cavity coherent amplitude starts from zero and gradually increases during the ring-up step, allowing the qubit to follow the cavity dynamics adiabatically~\cite{Dumas2024}. 
Therefore, to mimic our protocol using the displaced JC model, the preferable qubit initial state can be the lower-energy eigenstate of the Hamiltonian of the qubit subsystem in Eq.~\eqref{eq:H_JC_D}, which can be expressed as
\begin{equation}
    \ket{Q} = \hat R_y(\theta/2)\ket{g}.
\end{equation}
Namely, the composite initial state on the basis for the Hamiltonian in Eq.~\eqref{eq:H_JC_RD_main} can be represented as
\begin{equation}
    \ket{\Psi^{\text{RD}}} = \hat R_y^{\dag}(\theta/2)\hat D^{\dag}_{\hat a}(\alpha_{\mathrm{cav}})\ket{\Psi} = \ket{P}\otimes\ket{g}.
    \label{eq:psi_D}
\end{equation}
With this formulation, simulating the dynamics $\ket{\Psi}$ under the Hamiltonian in Eq.~\eqref{eq:H_JC} is equivalent to simulating the dynamics of $\ket{\Psi^{\text{RD}}}$ under the Hamiltonian in Eq.~\eqref{eq:H_JC_RD_main}, while the latter method requires only a small Hilbert space.

To analytically investigate the dominant effects of this model, we apply the Schrieffer–Wolff (SW) perturbation theory to the Hamiltonian in Eq.~\eqref{eq:H_JC_RD_main} and derive the effective Hamiltonian up to second-order perturbation (see Appendix~\ref{App:stab} for the explicit derivation):
\begin{equation}
    \begin{aligned}
        \hat H_{\mathrm{JC}}^{\mathrm{RD}}/\hbar\approx& -\tilde\Delta\dfrac{\hat\sigma_z}{2} +\chi(n_{\mathrm{cav}})\left(\hat a^{\dag}\hat a+\dfrac{1}{2}\right)\hat\sigma_z \\
        &+  J(n_{\mathrm{cav}})\left(\hat a^{\dag2}+\hat a^2\right)\hat\sigma_z + g_z(\hat a^{\dag}+\hat a)\dfrac{\hat\sigma_z}{2},
    \label{eq:H_SW}
    \end{aligned}
\end{equation}
where the effective frequency shift is given by
\begin{equation}
    \begin{aligned}
        \chi(n_{\mathrm{cav}}) = \chi_0\left(1+\dfrac{n_{\mathrm{cav}}}{2n_{\text{crit}}}\right)\left(1+\dfrac{n_{\mathrm{cav}}}{n_{\text{crit}}}\right)^{-3/2},
    \end{aligned}
    \label{eq:chi}
\end{equation}
and the effective squeezing rate is given by
\begin{equation}
    \begin{aligned}
        J(n_{\mathrm{cav}}) = \dfrac{\chi_0}{4} \dfrac{n_{\mathrm{cav}}}{n_{\text{crit}}}\left(1+\dfrac{n_{\mathrm{cav}}}{n_{\text{crit}}}\right)^{-3/2}, 
    \end{aligned}
    \label{eq:J}
\end{equation}
with $\chi_0 = g^2/\Delta$ corresponding to a state-dependent frequency shift in the conventional dispersive regime~($n_{\mathrm{cav}} =0$).

\begin{figure}[tp!]
    \centering
    \includegraphics[width=8.5cm]{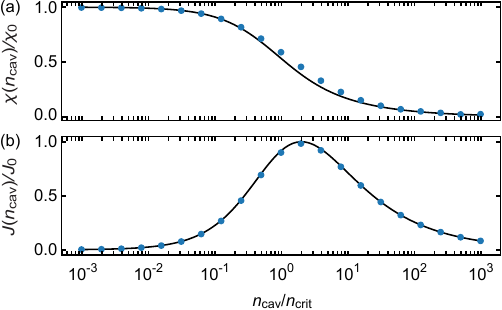}
    \caption{
    \textbf{}
    (a,~b) Effective frequency shift and squeezing rate as a function of the initial coherent cavity photon number in the displaced JC model. Both quantities are normalized by their analytical maximum values: $\chi_0 = g^2/\Delta$ and $J_0 = \chi_0/6\sqrt{3}$. The blue dots are the numerically simulated results, while the black solid lines correspond to the theoretical predictions. In the simulations, $\kappa/2\pi = \gamma/2\pi = 0$ are used.}
    \label{FIG-3}
\end{figure}

Since the qubit is in the lower eigenstate of $\hat\sigma_z$ in the new basis, and $\left[\hat\sigma_z,\hat H_{\mathrm{JC}}^{\mathrm{RD}}\right] = 0$, we can effectively replace $\hat\sigma_z$ with a classical number, i.e. $\hat\sigma_z\rightarrow-1$.
Therefore, the dominant effects on the cavity state are found to be a cavity frequency shift, squeezing, and coherent displacement.
Building on these insights, we numerically simulate the dynamics of the initial state of $\ket{\Psi^\text{RD}}$ under the Hamiltonian Eq.~\eqref{eq:H_JC_RD_main} using the method explained in Sec.~\ref{Sec:model}. In the simulation, $\ket{P}$ is chosen from the six Pauli eigenstates:
\begin{equation}
    \ket{P}\in \mathcal S_6\equiv \{\ket{0},\ket{1},\ket{\pm},\ket{\pm i}\},
    \label{eq:six_pauli}
\end{equation}
where $\ket{\pm} = (\ket{0}\pm\ket{1})/\sqrt{2}$, $\ket{\pm i} = (\ket{0}\pm i\ket{1})\sqrt{2}$, while $\ket{0}$ and $\ket{1}$ are the vacuum state and the single photon state of the cavity mode, respectively. 

We attribute the phase shift and the squeezing as the leading effects in cavity-state deformations, as illustrated in Fig.~\ref{FIG-2}a and suggested by the effective Hamiltonian Eq.~\eqref{eq:H_SW}.
We can therefore assume that the numerically simulated cavity state, centered around its coherent amplitude and denoted by $\hat{\rho}_\mathrm{sim}^D$, can be approximated by the following form:
\begin{equation}
    \hat{\rho}_\mathrm{sim}^D \approx \hat R_{\hat a}(\Delta\varphi)\hat S_{\hat a}(\xi)\hat \rho_\mathrm{sim}^\mathrm{C}\hat S^{\dag}_{\hat a}(\xi)\hat R^{\dag}_{\hat a}(\Delta\varphi), 
\end{equation}
where $\hat \rho_\mathrm{sim}^\mathrm{C}$ is the cavity quantum state after the correction of the phase shift and the squeezing.
$\hat S_{\hat a}(\xi) = \exp(\left(\xi\hat a^{\dag2}-\xi^{*2}\hat a^2\right)/2)$ is a cavity squeezing operator with squeezing parameter $\xi$ and $\hat R_{\hat a}(\Delta\varphi) = \exp(-i\Delta\varphi\hat a^{\dag}\hat a)$ is a cavity rotation operator with a phase shift of $\Delta\varphi$. 
In our analysis, we extract the squeezing parameter from the second-order moments of the squeezed quadrature $\hat x_{\mathrm{sq}}$ and the anti-squeezed quadrature $\hat x_{\mathrm{a.sq}}$ of the cavity field.
In order to extract the phase shift, we vary the phase rotation to maximize the overlap between the cavity quantum state $\rho^{D}_{\mathrm{sim}}$ and the initial state $\ket{P}$.
As described in Eq.~\eqref{eq:H_SW}, the phase shift and squeezing parameter of the cavity state are directly linked to the effective frequency shift and squeezing rate in the short-time evolution limit, giving rise to the following relationships:
\begin{equation}
    \Delta\varphi \approx -\chi t, ~\xi \approx 2iJt
    \label{eq:linear_fit}
\end{equation}
As a result, we extract the frequency shift $\chi$ and the squeezing rate $J$ from the numerical simulation results and compare them with the corresponding analytical predictions, as discussed below using Fig.~\ref{FIG-3}.

To quantify photon loss and heating effects beyond squeezing, we compute the change in the average photon number of the corrected state of $\hat \rho_\mathrm{sim}^\mathrm{C}$ relative to the initial state $\dyad{P}$:
\begin{equation}
\Delta n = \Tr[\hat{a}^\dagger \hat{a} \hat{\rho}_\mathrm{sim}^\mathrm{C}] - \Tr[\hat{a}^\dagger \hat{a} \dyad{P}],
\end{equation}
which serves as an additional indicator of the nonlinear dynamics of the system.
Finally, the state fidelity is computed as the maximum overlap between the initial state and the final cavity state centered on its coherent amplitude, by optimizing only the phase shift.
See Appendix~\ref{App:squeeze} for a detailed discussion of our method.


\begin{figure*}[tp!]
    \centering
    \includegraphics[width = 17cm]{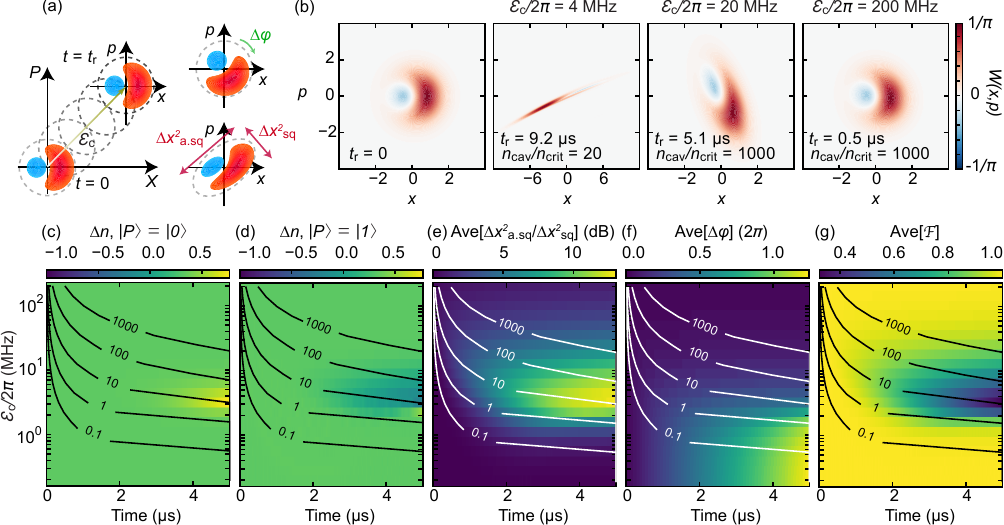}
    \caption{
    (a) Cavity-state dynamics under the driven JC Hamiltonian and illustration of the leading effects of cavity-state deformations: phase shift and squeezing. 
    (b) Wigner distribution of the cavity states centered around the coherent amplitudes after $t = 3~\mu\text{s}$ of evolution, shown for different cavity drive amplitudes. 
    (c,d) Photon number change $\Delta n$ for $\ket{P} = \ket{0}$ and $\ket{P} = \ket{1}$ as functions of time and cavity drive amplitude, respectively. 
    (e) Squeezing ratio $\Delta \hat{x}^2_{\text{a.sq}} / \Delta \hat{x}^2_{\text{sq}}$, averaged over $\ket{P} \in \{\ket{0}, \ket{1}\}$. 
    (f) Phase shift $\Delta\varphi$, averaged for $\ket{P} \in \{\ket{\pm}, \ket{\pm i}\}$. 
    (g) Fidelity of the cavity state to the initial state, averaged over $\ket{P} \in \mathcal{S}_6$. 
    The contour lines indicate the ratio of the coherent cavity photon number to the critical photon number.
    In the simulation, the cavity and the qubit are considered lossless, i.e., $\kappa/2\pi = \gamma/2\pi = 0$.
    }    
    \label{FIG-4}
\end{figure*}

In Fig.~\ref{FIG-2}b, we visualize the cavity states using Wigner distributions after an evolution time of $t = 3~\mu\text{s}$, for different initial coherent cavity photon numbers $n_{\mathrm{cav}}$, and compare them with the initial state at $t = 0$. More systematically, Figs.~\ref{FIG-2}c--g depict the time evolution of the aforementioned characteristic quantities for various initial coherent photon numbers $n_{\mathrm{cav}}$.
When the initial coherent cavity photon number is well below the critical photon number, only a finite phase shift emerges from the dispersive interaction with the qubit in the ground state (Fig.~\ref{FIG-2}f). 
As the value of $n_{\mathrm{cav}}$ approaches the critical photon number, the cavity–qubit interaction becomes strong, resulting in pronounced squeezing (Fig.~\ref{FIG-2}e) and substantial changes in the photon number that cannot be explained by squeezing alone (Figs.~\ref{FIG-2}c and d), indicating the presence of non-perturbative effects beyond the scope of perturbation theory.
When $n_{\mathrm{cav}}$ significantly exceeds the critical photon number, the cavity becomes effectively decoupled from the qubit, rendering both frequency shift and squeezing negligible. 
These trends are consistently reflected in the fidelity presented in Fig.~\ref{FIG-2}g.

In Fig.~\ref{FIG-3}, we quantitatively compare the analytical results with numerical simulations for the effective frequency shift and squeezing rate as functions of the initial coherent cavity photon number, averaged over $\ket{P} \in \{\ket{\pm}, \ket{\pm i}\}$. 
The analytical expressions of Eqs.~\eqref{eq:chi} and \eqref{eq:J} show excellent agreement with the simulation results across the entire range of coherent photon numbers. 
A pronounced peak in the squeezing rate appears when the coherent cavity photon number reaches twice the critical photon number, indicating that the system enters the \textit{critical regime}, where significant cavity-state deformations emerge. 
We define the \textit{dressed cavity regime} as the region where the cavity photon number is well below the critical photon number; in this regime, a finite frequency shift is observed while the squeezing remains negligible. 
Conversely, the \textit{bare cavity regime} corresponds to photon numbers far above the critical value, where both the frequency shift and squeezing are suppressed.
Based on these observations, the dressed, critical, and bare regimes can be characterized by the degree of cavity-state deformations, quantitatively given by the $g/\Delta$ ratio, or equivalently, by the critical photon number $n_{\text{crit}} = \Delta^2 / 4g^2$. 
\subsection{Resonantly driven JC model}
\label{Subsec: driven JC}

As shown in Fig.~\ref{FIG-4}a, we study possible perturbations on cavity quantum states being continuously displaced by the driven JC model. This mimics the ring-up step of our protocol, where similar cavity-state deformations, i.e., phase-shifting and squeezing, are expected to emerge. The Hamiltonian of the driven JC model is given by
\begin{equation}
    \hat H_{\text{JC,d}} /\hbar = \Delta\dfrac{\hat\sigma_z}{2} +  g(\ba^{\dag}\bs_-+\ba\bs_+)+\dfrac{\Ac}{2}(\hat a^{\dag}+\hat a),
\end{equation}
where the drive frequency is set to be on-resonance with the bare cavity frequency, and the drive amplitude is a real number for simplicity. To efficiently simulate the dynamics under this Hamiltonian, we adaptively displace the frame of the cavity following the method discussed in Sec.~\ref{Sec:model}. 


We simulate the full quantum dynamics of the driven JC model, in which the coherent cavity photon number is ramped up to a target value of $n_{\text{cav}} = 10^6$ using different drive amplitudes. 
In Fig.~\ref{FIG-4}b, we show the Wigner distributions of the initial cavity state and the states after time evolutions to achieve either maximal or targeted coherent photon numbers. 
In Figs.~\ref{FIG-4}c–g, we show the same quantities previously presented for the JC model in a displaced frame (Figs.~\ref{FIG-2}c–g), now plotted as functions of the evolution time and the cavity drive amplitude. The contours indicate the ratio of the coherent cavity photon number to the critical photon number. In this simulation, the trajectories of the cavity coherent amplitude for initial states $\ket{P}\in\{\ket{\pm},\ket{\pm i }\}$ are considered to follow the same trajectory as that of $\ket{P}\in\{\ket{0},\ket{1}\}$. 

When the drive amplitude is not sufficiently large, for example, $\Ac/2\pi = 4~\text{MHz}$, the cavity photon number reaches only up to $n_{\text{cav}}/n_{\text{crit}} \approx 20$ at its maximum. This limitation arises from the dispersive shift, which restricts the coherent cavity photon number from growing arbitrarily large. In such cases, the cavity quantum state undergoes strong deformations, as illustrated in the second panel of Fig.~\ref{FIG-4}b. In the case of a sufficiently strong cavity drive, on the other hand, the qubit is quickly saturated, resulting in the suppression of the dispersive shift, which leads to on-resonant driving of the cavity. Therefore, the coherent cavity photon number increases quadratically with the duration of the cavity drive. By comparing the third and fourth panels in Fig.~\ref{FIG-4}b, we observe that a stronger drive amplitude more effectively suppresses both the phase shift and squeezing when targeting the same coherent cavity photon number. These trends are quantitatively captured in Figs.~\ref{FIG-4}c–g.

\begin{figure}[tp!]
    \centering
    \includegraphics[width=8.5cm]{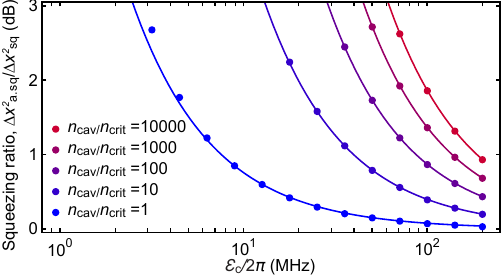}
    \caption{Cumulative squeezing ratio in resonantly driven JC model. The dots correspond to the numerically simulated results and the solid lines correspond to the analytical predictions, as given by Eq.~\eqref{eq:CSA_ana}.}
    \label{FIG-5}
\end{figure}

Building on the analytical results from the displaced JC model, as discussed in Sec.~\ref{Subsec: non-driven JC}, we seek an analytic solution for the cavity squeezing observed in the driven JC model.
Due to the adiabatic evolution of the qubit in the presence of a strong cavity drive, the cavity-qubit joint state can be approximated by Eq.~\eqref{eq:psi_D} during the entire ring-up step. Consequently, the squeezing rate at evolution time $t$ is given by Eq.~\eqref{eq:J} with coherent photon number $\ncav(t)$. Therefore, the cumulative squeezing parameter for the cavity quantum state can be estimated by integrating Eq.~\eqref{eq:J} over time:
\begin{equation}
    \begin{aligned}
        \abs{\xi(t_{\text r})} =& \abs{\int_0^{t_{\text r}}2iJ(\ncav(t))\mathrm{d}t} \\
               =& \int_0^{t_{\text r}} \dfrac{\abs{\chi_0}}{2} \dfrac{\ncav(t)}{\ncrit}\left(1+\dfrac{\ncav(t)}{\ncrit}\right)^{-3/2}\mathrm{d}t,
    \end{aligned}
    \label{eq:CSA}
\end{equation}
where $t_{\text r}$ is the final time of the ring-up step.
However, the coherent cavity photon number, $\ncav(t)$, is ramped up by the cavity drive ($\Ac$) in the presence of the photon-number-dependent dispersive shift $\chi(\ncav(t))$, as given by Eq.~\eqref{eq:chi}, hindering the analytical solution. Nevertheless, in the limit of a large cavity drive amplitude, the parameter regime of interest, the dispersive shift becomes negligible, facilitating a well-approximated analytical solution of $\ncav(t) \approx (\Ac t / 2)^2$. This allows for the full analytical expression of the cumulative squeezing parameter as
\begin{equation}
    \begin{aligned}
        \abs{\xi(t_{\text r})} & =\dfrac{\abs{\chi_0}\sqrt{n_{\mathrm{crit}}}}{\Ac}\int_0^{\bar{\alpha}_\mathrm{cav}} \bar{\alpha}^2\left(1+\bar{\alpha}^2\right)^{-3/2} \mathrm{d}\bar{\alpha}\\
       & = \dfrac{\abs{\chi_0}\sqrt{\ncrit}}{\Ac}\left[\ln\left(\sqrt{1+\dfrac{\ncav(t_\mathrm{r})}{\ncrit}} +  \sqrt{\dfrac{\ncav(t_\mathrm{r})}{\ncrit}}\right) \right.\\
       &\left.- \sqrt{\dfrac{\ncav(t_\mathrm{r})}{\ncav(t_\mathrm{r})+\ncrit}} \right],
    \end{aligned}
    \label{eq:CSA_ana}
\end{equation}
where the dimensionless parameter, given by $\bar{\alpha}_\mathrm{cav} = \sqrt{\ncav(t_\mathrm{r})/\ncrit}$, is linked to the final coherent cavity photon number.

\begin{figure}
    \centering
    \includegraphics[width=8.5cm]{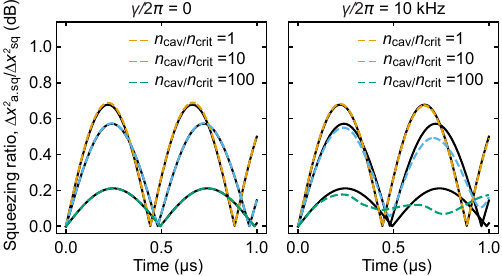}
    \caption{Time evolution of cavity squeezing ratio in off-resonantly driven JC model. The dashed lines correspond to the numerically simulated results, while the black solid lines correspond to the theoretical predictions, as given by Eq.\eqref{eq:r_transfer}. The cavity decay rate is not included in the simulation, i.e. $\kappa/2\pi = 0$.}
    \label{FIG-6}
\end{figure}

In Fig.~\ref{FIG-5}, we compare the numerical and analytical results for the cumulative squeezing ratio $\Delta\hat x_{\mathrm{a.s}}^2/\Delta\hat x_{\mathrm{s}}^2=\exp{4\abs{\xi(t_{\text r})}}$ as a function of the drive amplitude for different final coherent cavity photon numbers. The analytical predictions show good agreement with the numerical results in the large cavity drive limit.
We conclude that a stronger cavity drive decreases the duration during which the cavity quantum state traverses the critical regime, thereby minimizing the interaction time with the qubit and demonstrating the faithful transfer of the cavity quantum state from the dressed to the bare cavity regimes.
This can be easily confirmed by Eq.~\eqref{eq:CSA_ana}, as the cumulative squeezing parameter is inversely proportional to the cavity drive amplitude.

Moreover, the cumulative squeezing parameter can be further approximated in the limit of a large final coherent cavity photon number $\bar{\alpha}_{\mathrm{cav}}^2 = \ncav(t_\mathrm{r})/\ncrit\gg 1$, being expressed by
\begin{equation}
\begin{aligned}
        \abs{\xi(t_{\text r})}
            &\approx \dfrac{\abs{\chi_0}\sqrt{\ncrit}}{2\Ac}\left[ \ln\left(\dfrac{\ncav(t_\mathrm{r})}{\ncrit}\right)+2(\ln2 -1)\right].
    \label{eq:CSA_ana_simp}
\end{aligned}
\end{equation}
This analytical expression provides a clear guideline for selecting the system parameters to achieve a desired final photon number while minimizing detrimental cavity squeezing.




\subsection{Forced cavity oscillation in off-resonantly driven JC model}
\label{Subsec:forcedJC}

\begin{figure}[tp!]
    \centering
    \includegraphics[width = 8.5cm]{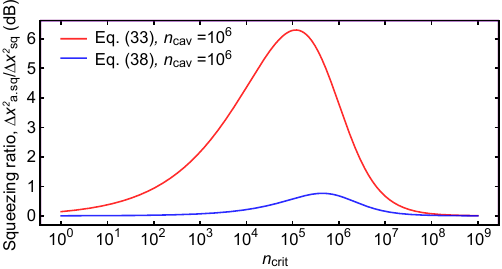}
    \caption{The cumulative squeezing ratios for the resonantly driven JC model (red) and the off-resonantly driven JC model (blue) are plotted as a function of the critical photon number $n_{\mathrm{crit}}$ for a fixed target cavity photon number $n_{\mathrm{cav}} = 10^6$ and $\chi_0/2\pi = -250$ kHz.}
    \label{FIG-7}
\end{figure}

As discussed in Sec.~\ref{Subsec: driven JC}, we confirm that the cumulative squeezing, identified as the leading source of unwanted cavity-state deformation during the ring-up step, can be effectively suppressed by a large cavity drive amplitude, enabling the cavity state to be transferred from the dressed-cavity regime to the bare-cavity regime with minimal distortion.
However, it remains nontrivial whether the residual finite squeezing in the bare-cavity regime could still accumulate during the transfer step.


To understand the evolution of the cavity squeezing during the transfer step, we consider the JC model with an off-resonant cavity drive, whose Hamiltonian is given by
\begin{equation}
    \hat H_{\text{JC,d}} /\hbar= \Deltac\hat{a}^\dagger \hat{a} +  \Deltaq\frac{\hat{\sigma}_z}{2} +  g(\ba^{\dag}\bs_-+\ba\bs_+)+\dfrac{\Ac}{2}(\hat a^{\dag}+\hat a),
    \label{eq:HJC_detuneddrive}
\end{equation}
where $\Deltac = \wc-\wdr$ and $\Deltaq = \wq - \wdr$.
For simplicity, the initial composite state is taken as a product state comprising a cavity coherent state, i.e., the vacuum state coherently displaced with photon number $n_\mathrm{cav} = \abs{\alpha_{\mathrm{cav}}}^2$ and a qubit state prepared identically to the method described in Sec.~\ref{Subsec: non-driven JC}.
For the transfer step, the cavity drive detuning is set to the mechanical frequency, i.e., $\Delta_{\text{c}}=\wm$. By performing SW perturbation theory, we can obtain the following effective cavity Hamiltonian:
\begin{equation}
    \begin{aligned}
        \hat H_{\mathrm{JC}}^{\mathrm{RD}}/\hbar \approx& \left(\Deltac -\chi(n_{\mathrm{cav}}) \right)\hat a^{\dag}\hat a -J(n_{\mathrm{cav}})\left(\hat a^{\dag2}+\hat a^2\right)\\
        &+\left(\dfrac{\Ac}{2}+\Deltac\alpha_{\mathrm{cav}}-\dfrac{g_z}{2}\right)(\hat a^{\dag}+\hat a),
    \label{eq:H_RD_off}
    \end{aligned}
\end{equation}
Here, we similarly replace $\hat\sigma_z$ by the classical number $-1$ and drop the qubit Hamiltonian and all the constant terms.
By forcing the drive amplitude to obey the following condition:
\begin{equation}
    \begin{aligned}
            \frac{\Ac}{2} =& -\Deltac \alpha_{\mathrm{cav}}  + \dfrac{g_z}{2},
    \end{aligned}
    \label{amp2}
\end{equation}
the net cavity drive term can be completely canceled, and the cavity state undergoes a forced oscillation at the drive frequency with a stationary coherent amplitude $\alpha_{\mathrm{cav}}$, thus effectively mimicking the dynamics during the transfer step of our protocol. (see Appendix~\ref{App:stab} for detailed calculations).

Using Eq.~\eqref{eq:H_RD_off} with the condition in Eq.~\eqref{amp2}, the time evolution of the effective cavity squeezing can be analytically solved (see Appendix~\ref{App:off_JC} for detailed derivation)
\begin{equation}
    \begin{aligned}
        \abs{\xi(t)} \approx &\abs{\dfrac{2J(\ncav)}{\Deltac-\chi(\ncav)}}\\
        &\times\abs{\sin\left(\sqrt{(\Deltac-\chi(\ncav))^2 -4\abs{J(\ncav)}^2}~t\right)}.
    \label{eq:r_transfer}
    \end{aligned}
\end{equation}
In our protocol, we can assume $\abs{J(\ncav)}, \abs{\chi(\ncav)} \ll\abs{\Deltac}$, which can be achieved by bringing the cavity to the bare regime where the coherent cavity photon number is larger than the critical photon number, as explained by Eq.~\eqref{eq:J}. In this limit, the expression of the squeezing parameter can be simplified as
\begin{equation}
    \abs{\xi(t)} \approx \frac{\abs{\chi_0}}{2\Deltac}\sqrt{\frac{n_{\mathrm{crit}}}{\ncav}}\abs{\sin(\Deltac t)}.
    \label{eq:r_transfer_lim}
\end{equation}

In Fig.~\ref{FIG-6}, we present both numerical and analytical results for the time evolution of the squeezing ratio of the cavity quantum state, evaluated for various initial coherent cavity photon numbers and qubit decay rates.
In these simulations, the cavity drive detuning is set to $\Deltac/2\pi = \wm/2\pi=1$ MHz. 
The squeezing ratio exhibits oscillations at a frequency determined by the detuning between the bare cavity frequency and the drive frequency without accumulations, while the analytical results perfectly predict these behaviors for the case of no qubit decay. Notably, the amplitude of these oscillations diminishes when the ratio $\ncav/\ncrit$ becomes large, a condition achieved when the coherent cavity photon number significantly exceeds the critical photon number at the beginning of the transfer step. This regime is favorable for performing optomechanical state transfer. Furthermore, we observe that qubit decay introduces slight perturbations to the squeezing oscillation but does not alter its overall oscillatory behavior.

\begin{figure*}[htp!]
    \centering
    \includegraphics[width=\textwidth]{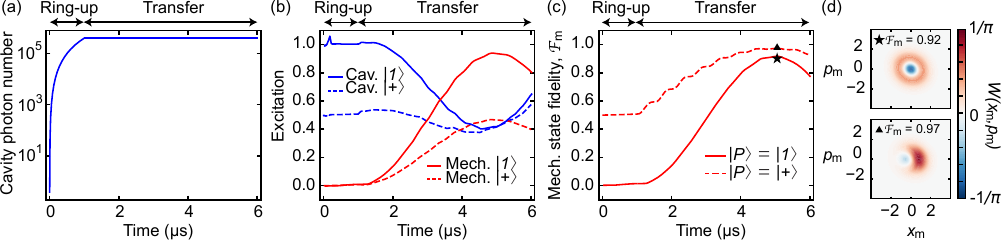}
    \caption{Optomechanical non-Gaussian-state transfer. 
    (a) Coherent cavity photon number. 
    (b) Excitation of cavity (blue) and mechanical oscillator (red) around their coherent amplitudes when the cavity is prepared in different non-Gaussian states. 
    (c) Mechanical state fidelity to the target non-Gaussian states $\ket{P}= |1\rangle$ and $|+\rangle$. 
    (d) Wigner function of the transferred mechanical state after a half cycle of energy exchange. The fidelity is optimized by correcting the phase of the state. In this simulation, $\kappa/2\pi = 1$ kHz, $\gamma/2\pi = 10$ kHz, while $\mathcal{E}_1/2\pi = 200$ MHz for the ring-up step.}
    \label{FIG-8} 
\end{figure*}

\subsection{Trade-off of system parameters in design strategy}
For optimal operation of our protocol, the system parameters must be optimized to simultaneously achieve controllability of non-Gaussian cavity states and strong suppression of unwanted squeezing effects.
However, a trade-off emerges regarding the state-dependent dispersive shift $\chi_0$.
The dispersive shift $\chi_0$ sets the time scale for non-Gaussian quantum control of cavity states using the qubit and needs to be sufficiently large to realize fast state control.
Meanwhile, the cavity squeezing parameters during both the ring-up and state-transfer steps, given by Eq.~\eqref{eq:CSA_ana} and Eq.~\eqref{eq:r_transfer}, respectively, increase linearly with the shift, i.e., with $\chi_0$.

Given a dispersive shift $\chi_0$, we can optimize the critical photon number $n_{\mathrm{crit}}$ to minimize squeezing effects, thereby determining the key system parameters $g$ and $\Delta$.
In Fig.~\ref{FIG-7}, we plot the squeezing ratios given by Eq.~\eqref{eq:CSA_ana} and the maximum of Eq.~\eqref{eq:r_transfer} as a function of the critical photon number $n_{\mathrm{crit}}$, fixing $\chi_0/2\pi = -250~\mathrm{kHz}$.
Here, $\Ac/2\pi = 200~\mathrm{MHz}$ during the ring-up step, $\Delta_\mathrm{c}/2\pi = \Omega_\mathrm{m}/2\pi = 1~\mathrm{MHz}$ during the state transfer step, and the target coherent cavity photon number of $\ncav=10^6$ are used, as these are primarily determined by other physical limitations.

We find that both squeezing effects are maximized when $n_{\mathrm{crit}}$ is set to be comparable with $\ncav$.
Although increasing $n_{\mathrm{crit}}$ above the target photon number can suppress squeezing, this requires both unrealistically large $g$ and $\Delta$.
Therefore, operating in the regime of a much smaller $n_{\mathrm{crit}}$, equivalently, a large hybridization ratio $g / \Delta$, is preferable to minimize unwanted squeezing.
We set $n_{\mathrm{crit}} = 100$ for a target coherent cavity photon number of $n_{\mathrm{cav}}\approx 10^6$, which balances squeezing suppression with fast non-Gaussian cavity-state generation under feasible system parameters.

\begin{figure*}[htp!]
    \centering
    \includegraphics[width = 17cm]{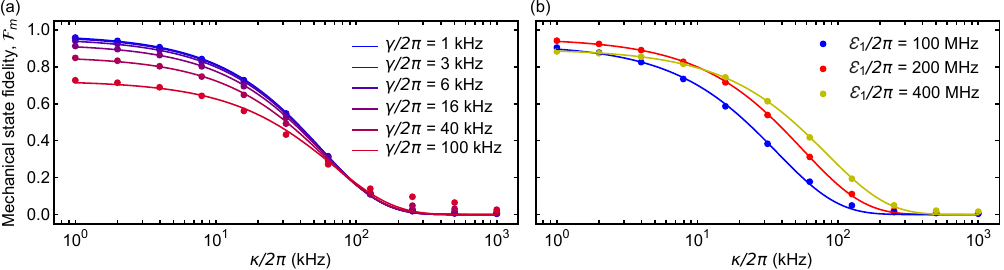}
    \caption{Dependence of the mechanical state fidelity $\mathcal{F}_m$ on system parameters, with the ring-up time fixed at $1~\mu\mathrm{s}$. (a) Mechanical state fidelity as a function of the cavity decay rate for different qubit decay rates, using a fixed ring-up cavity drive amplitude of $\mathcal{E}_1/2\pi = 200~\mathrm{MHz}$. (b) Mechanical state fidelity as a function of the cavity decay rate for different ring-up drive amplitudes, using a fixed qubit decay rate of $\gamma/2\pi = 10~\mathrm{kHz}$. The dots indicate the numerical simulation results, while the solid lines represent the exponential fits.}
    \label{FIG-9}
\end{figure*}
\section{Optomechanical state transfer}
\label{Sec:transfer}

We now return to the discussion of optomechanical state transfer in the full hybrid system, whose Hamiltonian is given by Eq.~\eqref{eq:total_H}. 
As described in Sec.~\ref{Sec:model}, our protocol begins with an on-resonant cavity drive to build up a sufficiently large coherent cavity photon number. The subsequent transfer step is initiated by abruptly switching the drive to a red-detuned frequency, thereby forcing the cavity into a steady oscillation and establishing a stationary optomechanical interaction.
In the limit of a lossless cavity, the drive amplitude after the switching must be properly configured to avoid possible transient dynamics. 
This approach stands in stark contrast to conventional cavity optomechanics, where cavity dynamics are typically much faster than those of optomechanical dynamics, allowing the cavity to immediately follow a quasi-steady state determined by an incident cavity drive field~\cite{Palomaki2013e,Delaney2019,Youssefi2023}.

In Fig.~\ref{FIG-8}, we present full simulations of non-Gaussian state transfer from the cavity to the mechanical oscillator, incorporating the decay rates for the cavity and qubit: $\kappa/2\pi = 1~\mathrm{kHz}$ and $\gamma/2\pi = 10~\mathrm{kHz}$. We prepare the cavity quantum state to be either in the single photon state $\ket{1}$ or in the superposition state $\ket{+}$. In Fig.~\ref{FIG-8}a, we show that the coherent cavity photon number $\ncav = |\alpha_{\text{cav}}|^2$ can be efficiently rung up to approximately $10^6$ within 1~$\mu$s using an on-resonant cavity drive with amplitude $\mathcal{E}_1/2\pi= 200$~MHz. Subsequently, the cavity coherent amplitude is stabilized by switching the drive frequency to be red-detuned from the cavity frequency by the mechanical frequency, with an optimal complex amplitude applied at the onset of the transfer step.
In contrast to the condition for the analytical model in Eq.~\eqref{amp2}, the optimal cavity drive amplitude is now obtained numerically as
\begin{equation}
    \begin{aligned}
            \frac{\mathcal{E}_2}{2} =& -\Deltac\expval{\hat a(t_{\mathrm{r}})} - g\expval{\hat\sigma_-(t_{\mathrm{r}})} \\
            &- g_0\expval{\hat a(t_{\mathrm{r}})}\left(\expval{\hat b^{\dag}(t_{\text r})}+\expval{\hat b(t_{\text r})}\right),
    \end{aligned}
    \label{amp2OM}
\end{equation}
where $\expval{\hat a(t_{\text r})}$ and $\expval{\hat b(t_{\text r})}$ correspond to the coherent amplitude of the cavity and the mechanical mode at the end of the ring-step step, while $g\expval{\hat\sigma_-(t_{\text r})}$ is computed numerically in the original qubit basis in the rotating frame of the drive frequency, serving as the analogue of the term $g_z/2$ in Eq.~\eqref{eq:r_transfer_lim}.

In Figs.~\ref{FIG-8}b and c, we demonstrate coherent energy exchange between the cavity and the mechanical oscillator, with the vacuum Rabi frequency given by the enhanced optomechanical coupling rate, i.e. $2g_{\text{OM}} = 2g_0\abs{\alpha_{\text{cav}}}$. The Wigner distributions of the mechanical states transferred from the cavity are shown in Fig.~\ref{FIG-8}d. These demonstrate that the non-Gaussian cavity states can be faithfully transferred to the mechanical oscillator with sufficiently high fidelity over 0.9. 

The reduction of the mechanical state fidelity is mainly attributed to the cavity decay rate $\kappa$. To enhance the fidelity, we can reduce the duration of the quantum state transfer by increasing the drive amplitude in the ring-up step, thereby increasing the coherent cavity photon number in the transfer step and the optomechanical coupling rate $g_{\text{OM}}$. Practically, the achievable coherent cavity photon number can be limited by cavity heating due to a strong cavity drive, which may introduce thermal noise in the cavity quantum states \cite{Teufel2011s, Youssefi2023}.

In Fig.~\ref{FIG-9}, we numerically illustrate the dependence of the achievable mechanical state fidelity on the cavity decay rate $\kappa$, the qubit decay rate $\gamma$, and the ring-up cavity drive amplitude $\mathcal{E}_1$. In the simulation, the ring-up time is fixed to be 1 $\mu$s for all individual points. In Fig.~\ref{FIG-9}a, the achievable mechanical state fidelity is shown for various qubit decay rates ranging from 1 kHz to 100 kHz and cavity decay rates ranging from 1 kHz to 1 MHz. As expected, the fidelity decreases exponentially with the cavity decay rate. Importantly, the fidelity shows a significant drop when the cavity decay rate becomes comparable to the vacuum Rabi oscillation frequency, which is given by the enhanced optomechanical coupling rate in the transfer step ($g_{\text{OM}}/2\pi \approx 100$ kHz). In contrast, no significant degradation is observed even when the qubit decay rate reaches a comparable magnitude. This behavior can be attributed to the saturation of the qubit by the end of the ring-up step, which mitigates the impact of qubit decay on the optomechanical state transfer. 

In Fig.~\ref{FIG-9}b, we compare the achievable mechanical state fidelity as a function of the cavity decay rate for different ring-up cavity drive amplitudes. Notably, in the limit of a small cavity decay rate, the fidelity slightly decreases at the highest drive amplitude $\mathcal{E}_1/2\pi = 400$ MHz. This reduction arises from an optomechanical interaction that off-resonantly couples the cavity to the mechanical oscillator during the ring-up step. 
Nevertheless, the resulting increase in the coherent photon number accelerates the optomechanical state transfer, leading to higher mechanical state fidelity in the cavity-decay-dominant regime. This observation highlights the existence of an optimal ring-up cavity drive amplitude, determined by the system parameters.

\section{Discussion and conclusion}

\label{Sec:discussion}
We developed a protocol to generate non-Gaussian states in a mechanical oscillator using a strongly driven cavity QED-optomechanical hybrid system. A non-Gaussian state is generated in the cavity by utilizing the nonlinearity of the cavity QED system in the dispersive regime, then coherently displaced from the dressed regime to the bare regime by applying a strong cavity drive, and finally transferred to the mechanical oscillator with a sufficiently strong optomechanical interaction enhanced by the coherent cavity photons. We numerically and analytically study possible cavity-state deformations induced in a strongly driven JC model, mainly characterized by effective phase shift and squeezing. Importantly, these undesired effects are significantly reduced by a sufficiently strong cavity drive that quickly brings the cavity state from the dressed cavity regime to the bare cavity regime, and they remain negligible during the transfer step. As a result, the sufficiently large optomechanical interaction together with the suppression of the unwanted cavity-state deformations guarantees the high-fidelity quantum-state transfer from the cavity to the mechanical oscillator, leading to the generation of a non-Gaussian mechanical state.

However, there are several points that need to be addressed to enhance the feasibility of our protocol. 
A complete description of the dipole–dipole interaction between the cavity and the qubit includes counter-rotating terms, which are typically neglected under the rotating wave approximation (RWA). 
One concern is that these counter-rotating terms may become significant in the presence of a strong cavity drive. 
To investigate this, we numerically simulate the full system dynamics, including the counter-rotating terms. 
Our results show that while these terms lead to a slight degradation in cavity-state fidelity during the ring-up step, the overall effect remains minimal. This confirms that our protocol is robust even when the counter-rotating terms are taken into account (see Appendix~\ref{App:RWA}).


Moreover, in circuit implementations of cavity QED, the qubit is typically realized by a superconducting transmon, which is inherently a multi-level system with weak anharmonicity, distinct from an idealized two-level system. In this case, the coupled system of a transmon and a microwave cavity is effectively described by a generalized JC Hamiltonian by replacing the Pauli operators of the standard JC Hamiltonian with bosonic ladder operators with a self-Kerr nonlinearity \cite{Blais2021}. To verify the applicability of our approach in a more practical setting, we apply our numerical method to a cavity QED system implemented with a transmon qubit (see Appendix~\ref{App:transmon} for details). Our result shows that the cavity-mechanics state transfer with a multi-level transmon is similar to the case of a two-level system, with negligible degradation in the final mechanical state fidelity. Nevertheless, recently reported results using a more rigorous model for the coupled system of a transmon and a cavity have shown that a strong drive on the transmon induces multi-photon transitions to coupled the qubit subspace to high-energy levels, resulting in transmon ionization, which may lead to degradation of qubit coherence and chaotic behaviors \cite{Shillito2022, Cohen2023}. A more complete study considering these effects will be followed up in the future for further experimental feasibility of our scheme.

\section*{Acknowledgment}
\label{Sec:acknowledgement}

This work has received funding from the European Research Council (ERC) under the EU H2020 research and innovation program, grant agreement No.~835329 (ExCOM-cCEO), as well as the Swiss National Science Foundation under grant agreement No.~231403 (CoolMe).

\appendix

\section{Benchmark of the simulation method: on-resonantly driven cavity}
\label{App:benchmark}

\begin{figure}[tp!]
    \centering
    \includegraphics[width=8.5 cm]{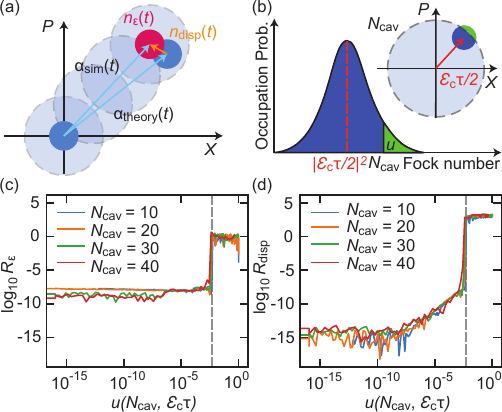}
    \caption{Benchmark of simulation method. (a) Schematic illustration of errors, including the residual photon number and the deviation of the coherent amplitude. (b) Photon number distribution of a coherent state with coherent amplitude $\Ac\tau/2$. The truncation error probability indicates the integral of the distribution above the size of the given Fock space, $N_{\mathrm{cav}}$ (green). (c) Residual photon number error rate and (d) coherent photon number error rate as a function of the truncation error probability $u$. The gray dashed lines correspond to the thresholds of the truncation error probability.}
    \label{FIG-10}
\end{figure}

We benchmark our numerical simulation method using an analytically solvable problem: an on-resonantly driven cavity (Fig.~\ref{FIG-10}a). 
For simplicity, the initial cavity state is set to the vacuum state, and cavity loss is neglected. 
Using the numerical method described in Sec.~\ref{Sec:model}, we simulate the dynamics in the rotating frame of the drive frequency by dividing the evolution into short intervals of duration $\tau$ and adaptively updating the displaced frame. 
This approach enables efficient computation of the time evolution of the cavity density operator, denoted by $\rho_\mathrm{sim}(t)$, in the original displaced basis after correcting for the displacement used during the simulation.


In theory, the vacuum state is displaced to a coherent state with a coherent amplitude corresponding to 
\begin{equation}
    \alpha_{\text{theory}}(t) = -i\Ac t/2,~t\in[0,T],
\end{equation}
where $T$ is the duration of the drive. To analyze how errors accumulate during the simulation, as illustrated in Fig.~\ref{FIG-10}a, we define the error rates per unit of coherent amplitude for both the residual and coherent photon numbers:
\begin{align}
    R_{\epsilon} \equiv& \max_{t\in[0,T]}\abs{\dfrac{\mathrm d n_{\epsilon}(t)}{\mathrm d \alpha_{\text {theory}}(t)}},\\
    R_{\text{disp}} \equiv& \max_{t\in[0,T]}\abs{\dfrac{\mathrm d n_{\text{disp}}(t)}{\mathrm d \alpha_{\text {theory}}(t)}},\\
\end{align}
where
\begin{align}
    n_{\epsilon}(t) =&\Tr[\rho_\mathrm{sim}^D(t) \hat a^{\dag}\hat a], \\
    n_{\text{disp}}(t) =&\abs{\alpha_{\text{sim}}(t)-\alpha_{\text{theory}}(t)}^2.
\end{align}
Here, we define $\alpha_{\text{sim}}(t) = \mathrm{Tr}[\hat{a} \hat{\rho}_\mathrm{sim}(t)]$ as the cavity coherent amplitude obtained from our numerical simulation. 
In addition, we define $\rho_\mathrm{sim}^D(t) = \hat{D}^\dagger(\alpha_{\text{sim}}(t)) \hat{\rho}_\mathrm{sim}(t) \hat{D}(\alpha_{\text{sim}}(t))$ as the cavity state centered around the coherent amplitude $\alpha_{\text{sim}}(t)$. 
Ideally, this cavity state corresponds to the vacuum state, resulting in $n_{\epsilon}(t)=0$.

The error rates are evaluated as a function of the truncation error probability $u$, defined by
\begin{equation}
    u(N_{\text{cav}}, \Ac\tau) \equiv \sum_{k = \lfloor N_{\text{cav}} \rfloor}^{\infty} 
    \frac{e^{-|\Ac \tau / 2|^2} |\Ac \tau / 2|^{2k}}{k!},
    \label{eq:u}
\end{equation}
where $N_\mathrm{cav}$ is the size of the Hilbert space used in the numerical simulation.
This error probability quantifies the photon number occupation probability falling outside the truncated Hilbert space during each interval of the coherent displacement dynamics.
As illustrated in Fig.~\ref{FIG-10}b, the truncation error probability captures the deviation between a perfect coherent state and its truncated approximation within a finite Hilbert space of dimension $N_{\text{cav}}$.
Intuitively, when the displacement $|\Ac \tau / 2|$ in each interval is much smaller than $\sqrt{N_{\text{cav}}}$, the value of $u$ becomes negligible, as the summation in Eq.~\eqref{eq:u} only accounts for a small tail of the Poisson distribution. As a result, the simulation errors introduced in each interval can be significantly suppressed.

In Figs.~\ref{FIG-10}c and d, we present the error rates $R_{\epsilon}$ and $R_{\text{disp}}$ as a function of the truncation error probability $u$. The length of each short interval $\tau$ is fixed to be 1 ns, and the displacement amplitude in each short interval is varied by changing the drive amplitude. We find that both errors are significantly low as long as $u \lessapprox 10^{-2.5}$. 
This threshold value can help us to determine $N_{\text{cav}}$ and $\tau$ with the given drive amplitude $\Ac$.

\section{Perturbation theory on JC model}
\label{App:stab}

In this section, we derive the effective Hamiltonian of the JC model in the displaced frame by applying perturbation theory. To be general, we consider the JC Hamiltonian with a cavity drive, given by $\hat H_{\mathrm{JC,d}}$ in Eq.~\eqref{eq:HJC_detuneddrive}. 
As mentioned in the main text, we first go to the displaced frame of $\alpha_0$ to minimize the required Fock space in the simulation. The Hamiltonian then becomes:
\begin{equation}
    \begin{aligned}
        \hat H^{\text{D}}_{\text{JC,d}}/\hbar =& \hat D^{\dag}(\alpha_{\mathrm{cav}})\hat H_{\text{JC,d}}\hat D(\alpha_{\mathrm{cav}})/\hbar \\
        =&\Deltac\hat a^{\dag}\hat a+\Deltaq\dfrac{\hat\sigma_z}{2}+  g\left(\hat a^{\dag}\hat\sigma_-+\hat a\hat\sigma_+\right) +  g\alpha_{\mathrm{cav}}\hat\sigma_x\\
        &+ \left(\dfrac{\Ac}{2} + \Deltac\alpha_{\mathrm{cav}}\right)\left(\hat a^{\dag} + \hat a\right),
    \end{aligned}
    \label{eq:H_RD_0}
\end{equation}
where we assume $\alpha_{\mathrm{cav}}$ and $\Ac$ to be real numbers for simplicity. We can also assume the frequencies follow the hierarchy: $\omega_{\mathrm{cav}}>\omega_{\mathrm{d}}>\omega_{\mathrm{q}}$. We thus have $\Deltac>0$ and $\Deltaq<0$.
Since the qubit can be approximated as remaining in the eigenstate of the Hamiltonian of the qubit subsystem, it is useful to diagonalize the qubit Hamiltonian by moving to the optimal qubit rotation basis, leading to
\begin{equation}
    \begin{aligned}
        \hat H^{\text{RD}}_{\text{JC}}/\hbar =& \hat R_y^{\dag}({\theta}/{2})\hat H^{\text D}_{\text{JC}}\hat R_y({\theta}/{2})/\hbar \\
        =&\Deltac\hat a^{\dag}\hat a - \tilde\Delta_{\mathrm q}\dfrac{\hat\sigma_z}{2}\\
        &+  g_1\left(\hat a^{\dag}\hat\sigma_-+\hat a\hat\sigma_+\right)+  g_2\left(\hat a^{\dag}\hat\sigma_++\hat a\hat\sigma_-\right)\\
        &+ \left(\dfrac{\Ac}{2} + \Deltac\alpha_{\mathrm{cav}} + \frac{g_z}{2} \hat\sigma_z\right)\left(\hat a^{\dag} + \hat a\right),
    \end{aligned}
    \label{eq:H_RD}
\end{equation}
where the parameters used here are defined in the main text of Sec.~\ref{Subsec: non-driven JC}, except for $\theta = \arctan(2g\alpha_{\mathrm{cav}}/\Deltaq)$ and $\tilde\Delta_{\mathrm{q}} =  \sqrt{\Deltaq^2 + 4g^2\abs{\alpha_{\mathrm{cav}}}^2}$ for greater generality.

We will then study the effective cavity-state deformation induced by Hamiltonian Eq.~\eqref{eq:H_RD} by performing perturbation theory.
Note that the last term in Eq.~\eqref{eq:H_RD} forms an effective cavity drive, which does not satisfy the condition for the perturbation calculation process. However, since the qubit is always in the ground state of the Hamiltonian in this frame, this term only displaces the cavity state. Thereby, we can neglect these terms in the following perturbation analysis.
We now separate the Hamiltonian Eq.~\eqref{eq:H_RD} into the diagonal part $\hat H_0$ and the perturbative off-diagonal part $\hat V$:
\begin{align}
    \hat H_0/\hbar = & \Deltac\hat a^{\dag}\hat a-\tilde\Delta_{\mathrm{q}}\dfrac{\hat\sigma_z}{2},\\
    \hat V/\hbar = & g_1\left(\hat a^{\dag}\hat\sigma_-+\hat a\hat\sigma_+\right) + g_2\left(\hat a^{\dag}\hat\sigma_++\hat a\hat\sigma_-\right).
\end{align}
By applying the SW-transformation and keeping the terms up to the second order, the Hamiltonian becomes:
\begin{equation}
    \begin{aligned}
        \hat H_{\text{eff}} =& e^{-\hat W}\hat H^{\text{RD}}_{\text{JC}}e^{\hat W}\\
        \approx& \hat H_{\text{eff}}^{(0)}+ \hat H_{\text{eff}}^{(1)}+\hat H_{\text{eff}}^{(2)},
    \end{aligned}
    \label{eq:Heff}
\end{equation}
where $\exp(-\hat W)$ is a unitary transformation and the zeroth-order effective Hamiltonian $\hat H_{\text{eff}}^{(0)}$ corresponds to the diagonal part $\hat H_0$. The first-order effective Hamiltonian is forced to be zero, i.e.
\begin{align}
    \hat H_{\text{eff}}^{(1)} =& [\hat H_0,\hat W] + V = 0.
\end{align}
Based on this condition, we can calculate the unitary operator, which is given by the following anti-Hermitian operator:
\begin{equation}
    \begin{aligned}
        \hat W =-\dfrac{g_1}{\Deltac+\tilde\Delta_{\mathrm{q}}}(\hat a^{\dag}\hat\sigma_--\hat a\hat\sigma_+)-\dfrac{g_2}{\Deltac-\tilde\Delta_{\mathrm{q}}}(\hat a^{\dag}\hat\sigma_+-\hat a\hat\sigma_-).
        \label{eq:W}
    \end{aligned}
\end{equation}
The second-order perturbation can be calculated as follows:
\begin{equation}
    \begin{aligned}
        \hat H_{\text{eff}}^{(2)}/\hbar=&\dfrac{1}{2}[\hat V, \hat W]/\hbar\\
        =& \left(-\dfrac{g_1^2}{\Deltac+\tilde\Delta_{\mathrm{q}}}+\dfrac{g_2^2}{\Deltac-\tilde\Delta_{\mathrm{q}}}\right)\left(\hat a^{\dag}\hat a+\dfrac{1}{2}\right)\hat\sigma_z\\
        &+\dfrac{1}{2}\left(\dfrac{g_1g_2}{\Deltac-\tilde\Delta_{\mathrm{q}}}-\dfrac{g_1g_2}{\Deltac+\tilde\Delta_{\mathrm{q}}}\right)\left(\hat a^{\dag2}+\hat a^2\right)\hat\sigma_z.
    \end{aligned}
    \label{eq:H_eff_2}
\end{equation}
The full effective Hamiltonian can be derived by bringing back the cavity drive terms, leading to
\begin{equation}
    \begin{aligned}
        \hat H_{\mathrm{JC}}^{\mathrm{RD}}/\hbar \approx&\Deltac\hat a^{\dag}\hat a - \tilde\Delta_{\mathrm{q}}\dfrac{\hat\sigma_z}{2}\\
        &+ \left(-\dfrac{g_1^2}{\Deltac+\tilde\Delta_{\mathrm{q}}}+\dfrac{g_2^2}{\Deltac-\tilde\Delta_{\mathrm{q}}}\right)\left(\hat a^{\dag}\hat a+\dfrac{1}{2}\right)\hat\sigma_z\\
        &+\dfrac{1}{2}\left(\dfrac{g_1g_2}{\Deltac-\tilde\Delta_{\mathrm{q}}}-\dfrac{g_1g_2}{\Deltac+\tilde\Delta_{\mathrm{q}}}\right)\left(\hat a^{\dag2}+\hat a^2\right)\hat\sigma_z\\
        &+\left(\dfrac{\Ac}{2}+\Deltac\alpha_{\mathrm{cav}}+\dfrac{g_z}{2}\hat\sigma_z\right)\left(\hat a^{\dag}+\hat a\right).
    \end{aligned}
\end{equation}
Furthermore, the frequency of the rotating frame is always close to the cavity frequency, allowing $\tilde\Delta_{\mathrm{q}} \approx \sqrt{\Delta^2 + 4g^2\abs{\alpha_{\mathrm{cav}}}^2}$. As a result, the effective Hamiltonian can be approximated as
\begin{equation}
    \begin{aligned}
        \hat H_{\mathrm{JC}}^{\mathrm{RD}}/\hbar \approx& \Deltac\hat a^{\dag}\hat a - \tilde\Delta_{\mathrm{q}}\dfrac{\hat\sigma_z}{2}\\
        &+\chi(n_{\mathrm{cav}}) \left(\hat a^{\dag}\hat a+\dfrac{1}{2}\right)\hat\sigma_z +J(n_{\mathrm{cav}})\left(\hat a^{\dag2}+\hat a^2\right)\hat\sigma_z\\
        &+\left(\dfrac{\Ac}{2}+\Deltac\alpha_{\mathrm{cav}}-\dfrac{g_z}{2}\right)\left(\hat a^{\dag}+\hat a\right).
    \end{aligned}
    \label{eq:H_RD_app}
\end{equation}
By assuming the qubit stays at the lower eigenstate of $\hat\sigma_z$, we can replace the $\hat\sigma_z$ by a classical number $-1$. Under this approximation, the Hamiltonian Eq.~\eqref{eq:H_RD_app} effectively becomes Eq.~\eqref{eq:H_RD_off} in the main text. Furthermore, when the additional conditions $\Deltac = 0$ and $\Ac = 0$ are imposed, this Hamiltonian can be further simplified to Eq.~\eqref{eq:H_SW} in the main text.

\section{Effective cavity dynamics in Off-resonantly driven JC Hamiltonian}
\label{App:off_JC}

As explained in the main text, the effective dynamics of the cavity state centered on the forced oscillating coherent amplitude in the off-resonantly driven JC model is well captured by the off-resonant squeezing interaction. The Hamiltonian is simply given by
\begin{equation}
    \hat H_{\mathrm{cav}}^{\mathrm{eff}} /\hbar=\Deltac \hat a^{\dag}\hat a+  J\hat a^{\dag2}+J^*\hat a^2
    \label{eq:H_J}
\end{equation}
Here, we discuss the analytical solution for the cavity dynamics governed by this Hamiltonian.

We begin by finding the eigenstates of the Hamiltonian, which is facilitated by transforming to a basis diagonalized by the squeezing operator:
\begin{equation}
    \hat H_{\text J}/\hbar = \hat S^{\dag}_{\hat a}(\xi)\hat H_{\mathrm{cav}}^{\mathrm{eff}} \hat S_{\hat a}(\xi)/\hbar = \tilde\Delta_{\mathrm{c}}\hat a^{\dag}\hat a,
    \label{eq:H_S}
\end{equation}
where $\xi = r e^{i\phi}$ is the squeezing parameter with amplitude given by $r=-1/2\arctan 2\abs{J}/\Deltac$ and phase given by $\phi = \arg{J}$. The eigenfrequency is $\tilde\Delta_{\mathrm{c}} = \sqrt{\Deltac^2-4\abs{J}^2}$. We denote $\ket{\psi(t)}$ the cavity quantum state in the original frame and $\ket{\psi(t)}_S$ the cavity quantum state in the new frame determined by the squeezing transformation. The cavity initial state in the original frame is denoted by $\ket{P}$. In the new frame, the cavity initial state is expressed as
\begin{equation}
    \ket{\psi(0)}_{\text S} = \hat S^{\dag}_{\hat a}(\xi)\ket{\psi(0)} = \hat S^{\dag}_{\hat a}(\xi)\ket{P}.
\end{equation}
At time $t$, the cavity state can be expressed as
\begin{equation}
    \begin{aligned}
        \ket{\psi(t)}_{\text S} =&e^{-i\hat H_{\mathrm{J}}t}\ket{\psi(0)}_{\text S}\\
        =& e^{-i\hat H_{\mathrm{J}}t}\hat S^{\dag}_{\hat a}(\xi)\ket{\psi(0)}\\
        =&e^{-i\tilde\Delta_{\mathrm{c}} t\hat a^{\dag}\hat a}e^{(\xi^*\hat a^2-\xi\hat a^{\dag2})/2}e^{i\tilde\Delta_{\mathrm{c}} t\hat a^{\dag}\hat a}e^{-i\tilde\Delta_{\mathrm{c}} t\hat a^{\dag}\hat a}\ket{P}\\
        =&e^{(\xi^*e^{i2\tilde\Delta_{\mathrm{c}} t}\hat a^2-\xi e^{-2i\tilde\Delta_{\mathrm{c}} t}\hat a^{\dag2})/2} \hat R_{\hat a}(\tilde\Delta_{\mathrm{c}} t)\ket{P}\\
        =&\hat S(re^{i(\phi-2\tilde\Delta_{\mathrm{c}} t)})\hat R_{\hat a}(\tilde\Delta_{\mathrm{c}} t)\ket{P}.
    \end{aligned}
\end{equation}
where $\hat R_{\hat a}$ is the rotation operator defined in Sec.\ref{Subsec: non-driven JC}. By applying the anti-squeezing operator, we can obtain the cavity state at time $t$ in the original frame:
\begin{equation}
    \ket{\psi(t)} = \hat S_{\hat a}(re^{i\phi})\hat S_{\hat a}^{\dag}(re^{i(\phi-2\tilde\Delta_{\mathrm{c}} t)})\hat R_{\hat a}(\tilde\Delta_{\mathrm{c}} t)\ket{P}.
    \label{eq:psi_S}
\end{equation}
This expression contains the composition of two squeezing operations with different squeezing parameters, which can be calculated by using the representation of SU(1,1) algebra \cite{Fisher1984}. Generally, for two arbitrary complex numbers $\xi_1 = r_1e^{i\phi_1}$ and $\xi_2 = r_2e^{i\phi_2}$, the following identity holds:
\begin{equation}
    \hat S_{\hat a}(\xi_1)\hat S_{\hat a}(\xi_2) = \hat S_{\hat a}(\xi_3)\hat R_{\hat a}(-\theta(\xi_1,\xi_2)),
    \label{eq:squeeze_composition}
\end{equation}
$\theta$ and $\xi_3 = r_3e^{i\phi_3}$ are functions of $r_1,~r_2,~\phi_2,~\phi_2$, which satisfy the following equations:
\begin{align}
    &\theta = -\dfrac{1}{2}i\log(\dfrac{1+e^{i(\phi_1-\phi_2)}\tanh r_1\tanh r_2}{1+e^{i(\phi_2-\phi_1)}\tanh r_1\tanh r_2})\label{eq:squeeze_composition_1},\\
    &e^{i\phi_3}\tanh r_3 = \dfrac{e^{i\phi_1}\tanh r_1+e^{i\phi_2}\tanh r_2}{1+e^{i(\phi_2-\phi_1)}\tanh r_1 \tanh r_2 }.\label{eq:squeeze_composition_2}
\end{align}
For our case, we can choose $\phi = 0, \xi_1 = r, \phi_1 = 0,\xi_2 = -re^{-2i\tilde\Delta_{\mathrm{c}} t}=re^{i(2\tilde\Delta_{\mathrm{c}} t+\pi)},\phi_2 = 2\tilde\Delta_{\mathrm{c}} t+\pi$. Then, by applying the squeezing composition rule given by Eqs.~\eqref{eq:squeeze_composition}-\eqref{eq:squeeze_composition_2}, we can re-express the cavity state Eq.~\eqref{eq:psi_S} by 
\begin{equation}
    \ket{\psi(t)} = \hat S(\tilde\xi(r,\tilde\Delta_{\mathrm{c}},t))\hat R_{\hat a}(-\theta(r,\tilde\Delta_{\mathrm{c}},t)+\tilde\Delta_{\mathrm{c}} t)\ket{P},
\end{equation}
where
\begin{equation}
    \abs{\tilde\xi} = \arctanh\abs{\dfrac{\tanh r(1+e^{i\phi_2})}{1+e^{i\phi_2}\tanh^2r}}.
\end{equation}
The rotation operator $\hat R_{\hat a}(-\theta+\tilde\Delta_{\mathrm{c}} t)$ simply rotates the cavity quantum state and therefore will not change the squeezing ratio. As $\abs{2J}/\abs{\Deltac} = 2\abs{r}\ll1$, then $\tanh r \approx r$, $\arctanh r\approx r$. Thus, we can simplify the squeezing amplitude $\abs{\tilde\xi}$ by keeping the leading order of $r$:
\begin{equation}
\begin{aligned}
    \abs{\tilde\xi}\approx& \dfrac{2\abs{J}}{\abs{\Delta_{\mathrm{c}}}}\abs{\sin(\tilde\Delta_{\mathrm{c}} t)}.
\end{aligned}
\label{eq:xi_theory}
\end{equation}

\begin{figure}[tp!]
    \centering
    \includegraphics[width=8.5cm]{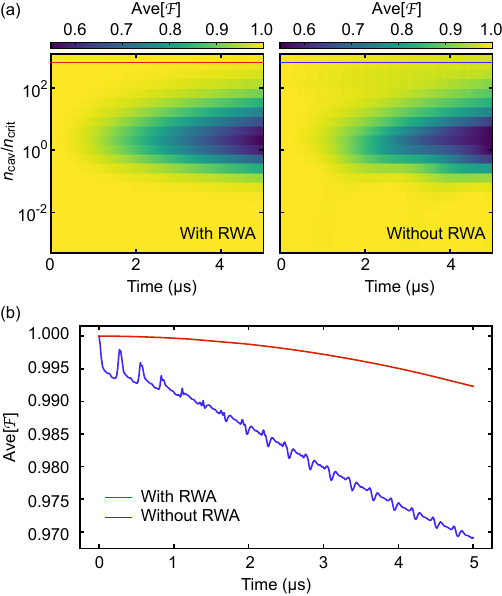}
    \caption{(a) Average cavity state fidelity over $\ket{P}\in\mathcal{S}_6$ for JC model in a displaced frame and Rabi model in a displaced frame. (b) Average cavity state fidelity at $n_{\mathrm{cav}}/n_{\text{crit}} = 10^3$ in the case with (orange) and without RWA (blue). $\kappa/2\pi = \gamma/2\pi = 0$ are used in this simulation.
    }
    \label{FIG-11}
\end{figure}

\section{Extraction of squeezing parameter and phase shift}
\label{App:squeeze}


Here, we discuss how to obtain the squeezing parameter from the simulated density matrix of the cavity, denoted by $\hat{\rho}_\mathrm{sim}$ in the original basis. 
To characterize the cavity state centered on the coherent amplitude, the following analysis is performed on the cavity state in the optimal displaced frame to remove the coherent amplitude, denoted by $\hat{\rho}_\mathrm{sim}^D$.
Here, the cavity state is obtained as $\hat{\rho}_\mathrm{sim}^D = \hat{D}^\dagger(\alpha_{\text{sim}}) \hat{\rho}_\mathrm{sim} \hat{D}(\alpha_{\text{sim}})$, where the coherent amplitude is determined as $\alpha_{\text{sim}} = \mathrm{Tr}[\hat{a} \hat{\rho}_\mathrm{sim}]$ using isotropic states, such as the vacuum and the single-photon states.

To find the squeezing angle, we first define rotated phase quadratures $\hat x_{\theta}$ and $\hat p_{\theta}$ by
\begin{align}
    \hat x_{\theta} =& \hat x\cos\theta +\hat p \sin\theta\label{eq:xtheta}\\
    \hat p_{\theta} =& -\hat x\sin\theta +\hat p \cos\theta,\label{eq:ptheta}
\end{align}
where $\hat x$ and $\hat p$ are the position and momentum quadratures of the cavity field in the displaced frame.
The rotation angle is determined such that $\hat x_{\theta}$ and $\hat p_{\theta}$ correspond to the squeezed and anti-squeezed quadrature basis, respectively. This results in 
\begin{equation}
    \expval{\hat x_{\theta}\hat p_{\theta} + \hat p_{\theta}\hat x_{\theta}} = 0.
\end{equation}
From this relation, the squeezing angle is determined as
\begin{equation}
    \theta = \dfrac{1}{2}\arctan\!\left(\dfrac{\expval{\hat x\hat p+\hat p\hat x}}{\expval{\hat x^2}-\expval{\hat p^2}}\right).
\end{equation}
Note that the angle can be uniquely determined such that $\expval{\hat x^2_{\theta}} \geq \expval{\hat p^2_{\theta}}$.
Using the quadratures rotated by this squeezing angle, the squeezing parameter $r$ and the residual photon number $n_\mathrm{res}$ can be obtained as 
\begin{equation}
    e^{2r} = \sqrt{\frac{\expval{\hat x^2_{\theta}}}{\expval{\hat p^2_{\theta}}}}. 
\end{equation}


To characterize the phase shift acquired during the dynamics under test, we can simply apply a rotation operator to the simulated cavity state $\rho^{D}_{\mathrm{sim}}$ and maximize its overlap with the initial cavity state.
Namely, we maximize the fidelity
\begin{equation}
\begin{aligned}
    F = \mel{\psi_0}{\hat R_{\hat a}(\beta)\,\rho^{D}_{\mathrm{sim}}\,\hat R^\dag_{\hat a}(\beta)}{\psi_0},
\end{aligned}
\label{eq:overlap}
\end{equation}
by varying the phase shift $\beta$.

By assuming the qubit remains in the lowest energy state, the evolution of the cavity state can be approximated with the effective Hamiltonian in Eq.~\eqref{eq:H_eff_2}, and decomposed as
\begin{equation}
    e^{-i\hat H^{(2)}_{\text{eff}}t/\hbar} \approx 
    \hat R_{\hat a}\!\left(-\chi t\right)\,
    \hat S_{\hat a}\!\left(i2Jt\right).
    \label{eq:evolutionH2eff}
\end{equation} 
Here, we omit the dependence on cavity photon number for the phase shift $\chi$ and the squeezing rate $J$. This approximation is valid when
\begin{equation}
    \abs{\chi J t^2}\ll 1.
\end{equation}
Moreover, the initial cavity states for our simulations are restricted to a superposition of the Fock states $\ket{0}$ and $\ket{1}$, i.e.,
\begin{equation}
    \ket{\psi_0} = c_0\ket{0}+c_1\ket{1}.
    \label{eq:psi0=0+1}
\end{equation}
The final cavity state can be approximated by $\rho^{D}_{\mathrm{sim}} \approx \ket{\psi^D_\mathrm{appox.}} \bra{\psi^D_\mathrm{appox.}}$, where $\ket{\psi^D_{\mathrm{approx.}}}$ is expressed as:
\begin{equation}
   \ket{\psi^D_{\mathrm{approx.}}} = \hat R_{\hat a}\!\left(-\chi t\right)\,
    \hat S_{\hat a}\!\left(i2Jt\right) \ket{\psi_0}
\end{equation}
Therefore, the approximated fidelity can be explicitly calculated as
\begin{equation}
    \begin{aligned}
        F \approx \abs{c_0}^4 S_{00}^2 + \abs{c_1}^4 S_{11}^2 + 
        2\,\abs{c_0}^2 \abs{c_1}^2 S_{00} S_{11}
        \cos\!\big(\chi t-\beta\big),
    \end{aligned}
\end{equation}
where $S_{mn}=\mel{m}{\hat S_{\hat a}(i2Jt)}{n}$.

As a result, maximizing the fidelity yields the optimal phase shift $\beta = \chi t$, allowing the extraction of the phase rotation imprinted on the cavity state.

\section{Validation of RWA}
\label{App:RWA}

In our model described in the main text, the coupling between the qubit and the cavity is given by the JC interaction as the counter-rotating terms are neglected by RWA. However, with the presence of a strong drive, the coupling between the JC ladders is enhanced by photon number, which may lead to the breakdown of RWA. Therefore, it is essential to consider the effects of the counter-rotating terms on cavity state deformation and the fidelity of the cavity state in our protocol.

Here, we consider the cavity dynamics in a displaced frame by taking the counter-rotating terms in the cavity-qubit coupled system, and study the validity of RWA in our protocol. In this case, the Hamiltonian in the frame of the bare cavity frequency becomes the Rabi Hamiltonian:
\begin{equation}
    \begin{aligned}
            \hat H_{\text{Rabi}}/\hbar =&\Delta\dfrac{\hat \sigma_z}{2} +  g(\hat a^{\dag}\hat\sigma_- + \hat a\hat\sigma_+) \\
            &+ g\left(\hat a^{\dag}\hat\sigma_+ e^{2i\wc t}+ \hat a\hat\sigma_- e^{-2i\wc t}\right),
    \end{aligned}
\end{equation}
In Figs.~\ref{FIG-11}a, we present the cavity state fidelity of the JC model and the Rabi model in the displaced frame, respectively. The cross sections across $n_{\mathrm{cav}}/n_{\text{crit}} = 1000$ for are shown in Fig.~\ref{FIG-11}b. The cavity state fidelity is averaged over different initial cavity states given by the condition Eq.~\eqref{eq:six_pauli}. To simulate the Rabi Hamiltonian, we adopt the same simulation approach by adaptively displacing the frame of the cavity, similar to what is described in Sec.~\ref{Sec:model}. We find that the counter-rotating terms suppress the cavity state fidelity, especially in the bare cavity regime. As shown in Fig.~\ref{FIG-11}c, the average fidelity decreases much faster in the case without RWA compared to the case with RWA, highlighting the substantial error accumulation induced by the counter-rotating terms. An oscillation in the average cavity state fidelity also emerges in the Rabi model without RWA, whereas no such oscillation is observed in the JC model with RWA applied. We attribute the reduction to the frequency mismatch of the qubit due to the additional frequency shift caused by the Bloch-Siegert effect and the oscillation to the higher-order harmonics \cite{Yan2015}. However, even in a high photon number, the counter-rotating terms do not cause a significant reduction to the cavity state fidelity and can therefore be neglected in the time scale of the quantum state transfer in our protocol.

\begin{figure}[tp!]
    \centering
    \includegraphics[width=8.5cm]{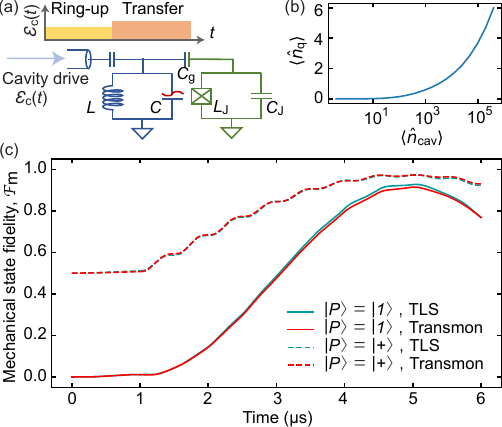}
    \caption{Implementation of hybrid system with transmon qubit. (a) Circuit diagram of the cavity QED-optomechanical hybrid system. The mechanical compliant resonator (blue) is coupled to a transmon qubit (gree). (b) Transmon excitation as a function of the coherent cavity photon number. (c) Comparison between the case of TLS (brown) and a multi-level transmon (red) with different initial cavity state. In this simulation, $K/2\pi = 200$ MHz, $\kappa/2\pi = 1$ kHz, $\gamma/2\pi = 10$ kHz, while $\Ac/2\pi = 200$ MHz for the ring-up step.
    }
    \label{FIG-12}
\end{figure}
\section{Implementation with transmon}
\label{App:transmon}

Our scheme can be implemented by integrating circuit optomechanics with a transmon qubit~\cite{Koch2007}. The corresponding circuit diagram is shown in Fig.~\ref{FIG-12}a. 
In the transmon regime, the qubit can be approximated as an anharmonic oscillator. Under the RWA, the coupled system is well described by an anharmonic oscillator coupled to a microwave cavity via a beam-splitter interaction. 
In the rotating frame of the cavity drive frequency, the Hamiltonian including the cavity drive term is given by
\begin{equation}
   \begin{aligned}
        \hat H_{\text T}/\hbar = &\Deltac\hat a^{\dag}\hat a+\Deltaq\hat q^{\dag}\hat q -\dfrac{K}{2}\hat q^{\dag2}\hat q^2+\wm\hat b^{\dag}\hat b\\
        &+  g(\ba^{\dag}\hat q+\ba\hat q^{\dag})+ g_0\ba^{\dag}\ba(\bb^{\dag}+\bb)+\dfrac{\Ac}{2}(\hat a^{\dag}+\hat a).
    \end{aligned}
    \label{eq:drivenHtransmon}
\end{equation}
Here, $K$ is the anharmonicity of the transmon qubit. To simulate the strongly driven coupled system, we adopt our approach based on adaptively displacing the transmon ladder operators to effectively minimize the required Hilbert space.

We first investigate the steady-state average transmon occupation under the dynamics governed by the Hamiltonian in Eq.~\eqref{eq:drivenHtransmon}, to ensure that the strong cavity drive used to increase the coherent photon number does not push the transmon population beyond the highest-energy bound state in the cosine potential. Such an excursion would cause the state to escape from the potential well, rendering our simplified transmon model invalid. 
To verify this, we simulate the transmon excitation as a function of the coherent cavity photon number, as shown in Fig.~\ref{FIG-12}b. 
Our results show that even under strong cavity driving—sufficient to induce the desired optomechanical coupling rate—the transmon excitation remains below the number of bound states, estimated as $\sim 2E_{\text{J}}/\sqrt{8E_{\text{J}}E_{\text{C}}}\approx\omega_{\mathrm{q}}/4K \approx 7$ for our system. 
This confirms the validity of our simplified transmon model in the relevant parameter regime.

Based on this consideration, we numerically simulate the state transfer between cavity and mechanical oscillator in a hybrid system based on the transmon and compare the result with the case when a two-level system is used as discussed in the main text. As illustrated in Fig.~\ref{FIG-12}, in both cases the non-Gaussian states $\ket{1}$ and $\ket{+}$ can be safely transferred to the mechanical oscillator with similar high fidelity, further confirming the feasibility of our scheme based on the circuit implementation.


\bibliography{ref_qubit_mechanics}

\end{document}